\documentclass[fleqn,usenatbib]{mnras}

\usepackage[T1]{fontenc}
\usepackage{ae,aecompl}

\usepackage{amstext}
\usepackage{graphicx}	% Including figure files
\usepackage{amsmath}	% Advanced maths commands
\usepackage{amssymb}	% Extra maths symbols
\usepackage{comment}
\usepackage{subfig}
\usepackage{float}
\usepackage{siunitx}
\usepackage{hyperref}
\usepackage{newtxtext,newtxmath}
\newcommand{\ds}{$\, {\rm deg}^2$}
\title[The GSMF from $0.1<z<2$]{Evolution of the galaxy stellar mass function: evidence for an increasing $M^*$ from $z=2$ to the present day}

\author[N. J. Adams et al.]{
N.J.~Adams$^{1}$,\thanks{E-mail: nathan.adams@physics.ox.ac.uk}
R.A.A.~Bowler$^{1}$,
M.J.~Jarvis$^{1,2}$,
B.~H{\"a}u\ss ler$^{3}$,
C.D.P.~Lagos$^{4,5,6}$
\\
$^{1}$Sub-department of Astrophysics, University of Oxford, Denys Wilkinson Building, Keble Road, Oxford OX1 2DL, UK.\\
$^{2}$Department of Physics, University of the Western Cape, Bellville 7535, South Africa \\
$^{3}$European Southern Observatory, Alonso de Cordova 3107, Vitacura, Santiago, Chile. \\
$^{4}$ICRAR, The University of Western Australia, 35 Stirling Highway, Crawley, WA 6009, Australia.\\
$^{5}$ARC Centre of Excellence for All Sky Astrophysics in 3 Dimensions (ASTRO 3D).\\
$^{6}$Cosmic Dawn Center (DAWN).
}

\date{Accepted XXX. Received YYY; in original form ZZZ}

\pubyear{2020}

\begin{document}
\label{firstpage}
\pagerange{\pageref{firstpage}--\pageref{lastpage}}
\maketitle

\begin{abstract}
Utilising optical and near-infrared broadband photometry covering $> 5\,{\rm deg}^2$ in two of the most well-studied extragalactic legacy fields (COSMOS and XMM-LSS), we measure the galaxy stellar mass function (GSMF) between $0.1 < z < 2.0$. We explore in detail the effect of two source extraction methods (SExtractor and ProFound) in addition to the inclusion/exclusion of \textit{Spitzer} IRAC 3.6 and 4.5$\mu$m photometry when measuring the GSMF. We find that including IRAC data reduces the number of massive ($\log_{10}(M/M_\odot) > 11.25$) galaxies found due to improved photometric redshift accuracy, but has little effect on the more numerous lower-mass galaxies.
We fit the resultant GSMFs with double Schechter functions down to $\log_{10}(M/M_\odot)$ = 7.75 (9.75) at z = 0.1 (2.0) and find that the choice of source extraction software has no significant effect on the derived best-fit parameters.
However, the choice of methodology used to correct for the Eddington bias has a larger impact on the high-mass end of the GSMF, which can partly explain the spread in derived $M^*$ values from previous studies. 
Using an empirical correction to model the intrinsic GSMF, we find evidence for an evolving characteristic stellar mass with $\delta \log_{10}(M^*/M_\odot)/\delta z$ = $-0.16\pm0.05 \, (-0.11\pm0.05)$, when using SExtractor (ProFound). 
We argue that with widely quenched star formation rates in massive galaxies at low redshift ($z<0.5$), additional growth via mergers is required in order to sustain such an evolution to a higher characteristic mass.
\end{abstract}

\begin{keywords}
galaxies: evolution -- galaxies: general -- galaxies: abundances %-- galaxies: high-redshift
\end{keywords}

%%%%%%%%%%%%%%%%%%%%%%%%%%%%%%%%%%%%%%%%%%%%%%%%%%

%%%%%%%%%%%%%%%%% BODY OF PAPER %%%%%%%%%%%%%%%%%%

\section{Introduction}

\subsection{The evolution of the galaxy stellar mass function}
The galaxy stellar mass function (GSMF) is a measurement of the cumulative effects of physical processes that enhance or hinder star formation within galaxies. These processes include merger events, internal feedback mechanisms (both supernova and active galactic nuclei (AGN) driven) and environmental effects. Understanding the balance between these processes is key to understanding how galaxies have been assembled over cosmic time.
Measurements of the local GSMF reveal a steep cut-off in the number of high-mass galaxies above a characteristic mass $\log_{10}(M^*/M_\odot)\sim 10.7$ (e.g.~\citealp{Baldry2012}). In addition to this, the population of very high-mass galaxies becomes increasingly quenched with time \citep[e.g.][]{Davidzon17,McLeod2020}.
Many theories have been proposed to explain why there is significant suppression in the star formation (often called quenching) of galaxies above this stellar mass, examples include but are not limited to: starvation/strangulation \citep{Larson1980,Kawata2008,McCarthy2008,Feldmann2011,Bahe2013,Feldmann2015}, virial shock heating \citep{Birnboim2003,Dekel2006,Cattaneo2006} and AGN feedback \citep{Binney1995,DiMatteo2005,Springel2005,Bower2006,Croton2006,Cattaneo2009,Fabian2012,Bongiorno2016,Beckmann2017}. 
To adequately test these theories and increase our understanding of how these processes influence galaxy evolution, simulations are required. Key milestones in testing theories such as these include being able to accurately reproduce the observed galaxy population through Luminosity/Mass Functions \citep[see][for recent reviews]{Somerville2015,Vogelsberger2020}. The evolving shape of the GSMF is dependent on all forms of stellar mass growth, including growth via merging events in addition to internal star formation \citep[e.g.][]{Rodriguez2015,Qu2017,Oleary2020}. Consequently, to increase our understanding of both quenching and merger rates, stellar mass functions have become a key benchmark for simulations in the past few years \citep{Henriques2013,Schaye2015,Pillepich2018,Lagos18} 
and hence, require observations to be reliable in order to tune and test these simulations.

The advance in deep extra-galactic surveys over the past decade has led to a better measurement of the GSMF at high-redshift.
Using predominantly photometric redshifts, several studies have now mapped out the evolution of the GSMF from $z = 0$--$3$ \citep[e.g.][]{Fontana2004,Perez2008,Marchesini2009,Pozzetti2010,Ilbert2013,Muzzin2013,Davidzon17,Wright18} and even up to $z\simeq 7$~\citep[e.g.][]{Verma2007,McLure2009,Stark2009,Grazain2015,Song2015,Thorne2020}.
Despite the increased cosmological volume probed by these studies, there is no consensus on the exact form of the GSMF over this epoch.
For example \citet{Wright18}, \citet{McLeod2020} and \citet{Thorne2020} find an approximately constant characteristic mass with redshift, but offset from each other by up to $M^* \sim 0.5\,$dex. Alongside this, other studies of high mass systems suggest merger events are required to explain the observed change in the size-mass relation of these objects \citep[e.g.][]{McLure2013}. The rate of mergers required would consequently influence the evolution of the shape of the stellar mass function at the high mass end.
To solve these issues, the precise number densities of galaxies with masses greater than $M^*$ are required.
Uncertainties in both the stellar mass and number density of these objects can have a much larger impact on the measured shape and characteristic mass of the derived GSMF than the more numerous low-mass sources. This is due to the need to quantify a steep exponential fall off, with relatively few galaxies, that suffer from higher levels of cosmic variance \citep[e.g.][]{Moster11}.
In this work we exploit deep, wide-area optical and IR imaging to measure the GSMF over $> 5\,{\rm deg}^2$, leveraging the large area to understand, in particular, the evolution in the number density of the most massive galaxies between $0.1<z<2.0$.

\subsection{Measuring accurate photometric redshifts and stellar masses}
In the coming decade, the Vera Rubin Observatory \citep{LSST} and \textit{Euclid} \citep{Euclid} survey programmes will be conducting broadband observations to further improve both the area and depths achieved within popular multi-wavelength fields such as COSMOS and XMM-LSS, among others. In parallel to these photometric surveys, a number of spectroscopic campaigns are also set to begin operations in the next couple of years. MOONS \citep{MOONS} and WAVES \citep{WAVES} are examples of multi-object spectrograph (MOS) surveys and both have an immediate need for high quality photometric redshift and physical parameter estimates in order to plan survey operations and develop final target catalogues ahead of commissioning. Consequently, there is demand for a maintained compilation of broadband photometry and photometric redshifts based on the improved optical and NIR data that has been obtained in recent years in order to best prepare for these upcoming projects \citep[past examples including:][]{Laigle16,Alarcon20}.

Moreover, with next-generation telescopes and survey programmes comes next-generation software and analysis pipelines. A workhorse in photometric source extraction for over 20 years has been {\tt SExtractor} \citep{Bertin1996}, but in recent years there has been a renewed effort in tackling the problem of obtaining accurate flux measurements from images with robust uncertainties and improved handling of source blending. One product of this effort has been {\tt ProFound} \citep{Robotham18}, which potentially provides improved photometry, particularly for extended sources, over a variety of wavelengths due to the non-parametric apertures used \citep[e.g.][]{Davies18,Hale19,Bellstedt20}. In this study we produce multiple catalogues of broadband photometry and using these different source extraction methods. The use of two source extraction tools, as well as varying the use of {\em Spitzer}/IRAC data,  allows for any potential bias' in the GSMF due to these different effects to be explored.

This paper is presented as follows: In Section 2 we describe the data used in producing our photometric catalogues. In Section 3 we describe the methods used in source extraction, fitting for photometric redshifts and derivation of basic physical properties of galaxies. In Section 4 we present the measured  GSMF in the redshift range $0.1 < z < 2.0$ and in Section 5 we present the results of modelling the GSMF, and compare the models with the results from previous studies. In section 6 we explore the time evolution of the GSMF and how our observations compare to results from simulations. 
We finally present our conclusions in Section 7. 
Throughout this paper we use the AB magnitude system \citep{Oke1974,Oke1983} and assume $\Lambda$CDM cosmology with $H_0=70$\,km\,s$^{-1}$\,Mpc$^{-1}$, $\Omega_{\rm M}=0.3$ and $\Omega_{\Lambda} = 0.7$. 

\section{Data} \label{sec:data}

The following sub-sections describe the data in the two fields used in the construction of our photometric catalogues and subsequent measurement of the GSMF.

\subsection{COSMOS}

The COSMOS field \citep{Scoville2007} is one of the most widely studied multi-wavelength fields in extra-galactic astrophysics, with data spanning the X-ray through to the radio domains over $\sim$ 2\ds \, of the sky centred on the J2000 coordinates of RA = 150.12\,deg (10:00:28.6) DEC = +2.21\,deg (+02:12:21.0). 

The imaging data over this field that are used in the construction of our catalogues is derived from four telescopes. The bluest coverage is from the Canada-France-Hawaii Telescope Legacy Survey \citep[CFHTLS;][]{Cuillandre2012} which has an ultra-deep pointing in the central square degree of the field, and we restrict our analysis to this area for this reason. From this survey we make use of the $u^*$-band. For the optical coverage we take data from the ultra-deep component of the HyperSuprimeCam Strategic Survey Programme DR1 \citep[HSC;][]{Aihara2014,Aihara2017}. Near-infrared imaging is sourced from the UltraVISTA survey \citep{McCracken2012}. We make use of the fourth data release of UltraVISTA (DR4) which has a tiered observing strategy, leading to a striped pattern of near-infrared coverage across the field. The `deep' component is approximately 1 magnitude shallower than the `ultra-deep' region. 3.6 and 4.5 micron photometry is obtained from the \textit{Spitzer} Extended Deep Survey \citep[SEDS;][]{Ashby2013}, \textit{Spitzer}
Matching Survey of the UltraVISTA ultra-deep Stripes survey \citep[SMUVS;][]{Ashby2018} and \textit{Spitzer} Large Area Survey with Hyper-Suprime-Cam \citep[SPLASH;][]{Capak2012}. The $5\sigma$ detection limits for the 2 (2.8) arcsecond photometry in each optical/NIR (\textit{Spitzer}) bands are outlined in Table~\ref{tab:FiveSig}. This table also highlights the impact of the tiered structure of the UltraVISTA survey.

\subsection{XMM-LSS}

With a much greater area (at $\simeq 4.5$ \ds), the XMM-Large Scale Structure field is one of 3 deep fields that make up the Vista Infrared Deep Extra-galactic Observations (VIDEO) survey \citep{Jarvis2013}. Located at RA = 35.5\,deg (02:22:00.0) DEC = -4.8\,deg (-04:48:00.0), our study focuses on regions of the field with high quality HSC observations. We mask parts of the edge of the field due to the lack of overlapping coverage between HSC and VISTA  \citep[see][for more information on overlapping coverage from the different surveys and variable depths]{Bowler20}. The total area of the field that is used after considering the overlap of telescope footprints is 4.23\ds\, (giving a total of 5.23\ds\, when combined with COSMOS).

We use the same photometric bands as those used in the COSMOS field. Unlike in the COSMOS field, the optical coverage in XMM-LSS is not uniform while the near-infrared is more uniform. 
$u$-band imaging over the full XMM-LSS field was obtained from the CFHTLS Wide survey in addition to the CFHTLS-D1 field which covers a 1\ds \, patch where $u^*$ observations are 1 magnitude deeper than the rest of the field \citep{Cuillandre2012}. HSC SSP covers a different 1.5\ds \, region of the field \citep[centred on UKIDSS UDS][]{Lawrence2007} which has `ultra-deep' coverage. Near-infrared photometry is derived from the final data release of VIDEO \citep{Jarvis2013} and \textit{Spitzer} data is sourced from the \textit{Spitzer} Extragalactic Representative Volume Survey \citep[SERVS;][]{Mauduit2012}. The depths of the images in each broadband filter are also outlined in Table~\ref{tab:FiveSig}. 

\begin{table}
    \centering
      \caption{Summary of the $5\sigma$ detection depths within the COSMOS and XMM-LSS fields. Depths are calculated in 2\arcsec ~diameter circular apertures (2.8\arcsec ~for IRAC bands due to \textit{Spitzer}'s larger point spread function) that were placed on empty regions of the image. Values are grouped into Deep (D) and Ultra-Deep (UD) sub-regions. 
      The XMM-LSS field has a single ultra-deep pointing from HSC centred on the UDS field and so a third of XMM-LSS has this deeper optical coverage. Similarly the $u^*$ coverage has a single square degree of ultra-deep coverage in a separate part of XMM-LSS as part of the CFHT Legacy Survey observing programme.}
    \label{tab:FiveSig}
    \begin{tabular}{cccccc}
    \hline
    Filter & COS-UD & COS-D & XMM-UD & XMM-D & Origin \\
    \hline
    $u^*$ & $ 27.2$& --  & $25.9$  &$25.9-26.9$& CFHT\\
    $g$ & $ 27.1$  & -- &$ 27.0$ &$ 26.5$& HSC\\
    $r$ & $ 26.7$  & -- &$ 26.5$ &$ 25.9$& HSC\\
    $i$ & $ 26.6$  & -- &$ 26.4$ &$ 25.5$& HSC\\
    $z$ & $ 25.9$  & -- &$ 25.7$ &$ 24.7$& HSC\\
    $y$ & $ 25.5$  & -- &$ 25.0$ &$ 24.1$& HSC\\
    $Y$ & $26.3$  &$ 25.2$ &$ 25.4$ & --& VISTA\\
    $J$ & $26.1$  &$ 25.0$ &$ 24.9$ & --& VISTA\\
    $H$ & $25.7$  &$ 24.6$ &$ 24.4$ & --& VISTA\\
    $K_{\rm s}$ & $25.4$  &$ 25.0$ & $24.0$ & -- & VISTA\\
    3.6 & $24.8$  & -- &$ 24.3$ & -- & Spitzer\\
    4.5 & $24.8$  & -- &$ 24.0$ & -- & Spitzer\\
    \hline
    \end{tabular}
\end{table}

\section{Photometric catalogues and derived products}

For our study we define and produce four separate stellar mass estimates for all galaxies identified in the near-infrared. The various stellar mass estimates are determined by using two source extraction algorithms and including/excluding the \textit{Spitzer}/IRAC 1 and 2 bands. The goal here is to examine the impact of these two variables in the methodology, such as how different measurements of total flux translates to differences in stellar-mass estimation and how the use of \textit{Spitzer}/IRAC bands affects photometric redshift performance and the final distributions in stellar mass.

We refer to these catalogues as `SExtractor/SE' and `ProFound/PF' when using them in a lengthened/shortened format. When the two \textit{Spitzer}/IRAC bands are included in any analysis, the suffix `+IRAC' is added to the catalogue name.

\subsection{SExtractor photometry}

Source finding is performed in SExtractor \citep{Bertin1996} with the $K_s$-band image used for object selection. The fiducial photometry was derived using 2 arcsecond diameter circular apertures placed at the location of sources found by SExtractor. As \textit{Spitzer} has a larger point spread function, for the IRAC bands we use 2.8 arcsecond diameter circular apertures \citep{Bowler20}. 
The optical and NIR photometry is aperture corrected using a point-spread-function (PSF) generated in each band by {\sc PSFEx} \citep{Bertin2011} based on cut-outs of point-sources. This is calculated separately for the COSMOS field and over each of the three VISTA/VIDEO tiles in XMM-LSS. For \emph{Spitzer} photometry, we use an aperture correction derived in the \emph{Spitzer} handbook\footnote{The Spitzer handbook can be found at: \url{https://irsa.ipac.caltech.edu/data/SPITZER/docs/spitzermission/missionoverview/spitzertelescopehandbook/}}.  Alongside this we also measure the \texttt{MAG\_AUTO} parameter from SExtractor to estimate the total flux for each object. The flexible aperture used in the measurement of \texttt{MAG\_AUTO} is generated from the $K_s$-band detection image. 
Due to the significantly larger PSF of the two \emph{Spitzer} bands, carrying over the same aperture from the $K_s$ band would lead to an underestimation of the total flux. 
To solve this problem, we derive a correction to translate aperture flux ($f$) to total flux ($F$). This correction follows the average value of $F_{\nu,IRAC} = f_{\nu,IRAC} \times \frac{F_{\nu,K_s}}{f_{\nu,K_s}}$ for objects in bins of $K_s$ magnitude and redshift used in the later calculation of the GSMF. The use of this correction is sanity checked against the methodology we use with the {\sc ProFound} catalogues and is found to agree well for luminous objects.
Errors on the SExtractor photometry are calculated based on local depth maps generated by inserting apertures in empty locations of the field \citep[the same method as applied in][]{Bowler20}.

\subsection{Profound photometry}
In addition to the SExtractor photometry, we produce {\sc ProFound} \citep{Robotham18} catalogues selected on a weighted stack of the VISTA-$Y$, $J$, $H$ and $K_{\rm s}$ bands\footnote{We note that this does not affect our results as we subsequently apply a conservative limiting magnitude cut based on the $K_{\rm s}$-band}. {\sc ProFound} generates two photometric measurements for each object, a `total' flux and a `colour' flux. {\sc ProFound} operates by iteratively dilating the aperture encompassing each galaxy in each image until it meets the local background. This results in a morphologically derived aperture around each identified object for each photometric band. For our purposes we only make use of the total flux measurements and the associated errors from the {\sc ProFound} output. {\sc ProFound} flux errors are calculated by combining the errors resulting from the sky root mean square (RMS), errors from the sky subtraction process and object shot noise. Because this dilation process is performed for each photometric band, the larger~\emph{Spitzer}/IRAC PSF is taken into account and thus no alterations are required to obtain total fluxes in these bands. 

To validate both the ProFound and SExtractor photometry. We compare our measured total photometry against measurements made in other catalogues which have targeted these fields. These comparison catalogues are the COSMOS2015 catalogue for the COSMOS field \citep{Laigle16} and the  Subaru/XMM-Newton Deep Field (SXDF) catalogue \citep{Mehta2018}. We find that the vast majority of our photometric bands are consistent ($\ll0.1$~mag median offsets) with the measurements from these reference catalogues. The methodology we use to determine total SExtractor \emph{Spitzer} fluxes for bright, resolved sources is also found to be consistent with these catalogues. The greatest difference in photometry is found when running ProFound on \emph{Spitzer} images in XMM-LSS. For bright sources ($m<20$), the photometry is consistent with our SExtractor catalogue and the reference catalogue. However, for fainter objects ($m>20$) the measured flux in ProFound is found to be fainter than measured with SExtractor (up to 0.2~mags offset at $m\sim24$). Similarly, colour distributions match those of the reference catalogues, with the exception of the aforementioned ProFound IRAC offsets in XMM-LSS.

\begin{table}
\caption{The list of zero-point offsets applied to the 2 arcsecond aperture photometry after optimising the template fitting process against a large spectroscopic sample. Offsets listed are in magnitudes and are calculated separately for each field.}
\centering
\begin{tabular}{l|ll}
\hline
Band                 & XMM Offset & COS Offset \\ \hline
$u^*$ & 0.138      & 0.136      \\
g                    & -0.062     & -0.066     \\
r                    & -0.096     & -0.078     \\
i                    & -0.011     & -0.026     \\
z                    & -0.048     & -0.037     \\
y                    & -0.064     & -0.054     \\
Y                    & 0.078      & -0.006     \\
J                    & 0.037      & 0.020      \\
H                    & 0.077      & 0.084      \\
Ks                   & 0.025      & 0.024      \\
IRAC1                & 0.006      & 0.005      \\
IRAC2                & -0.048     & 0.013      \\ \hline
\end{tabular}
\label{tab:zeropoint}
\end{table}

\begin{table*}
\caption{Summary of the photometric redshift statistics. Displayed are the outlier rates and Normalised Median Absolute Deviation (NMAD) of the catalogues when compared to a large spectroscopic sample. We cut show these numbers in cuts of i-band magnitudes to enable comparison with results from \citet{Laigle16}. The cut of $m_i < 22$ corresponds to the brightest 50 per cent of the spectroscopic sample we compare to in COSMOS and the brightest 60 per cent in XMM-LSS.}
\label{tab:stats}
\begin{tabular}{l|llll}
        & Outlier Rate (\%) $m_i<22$ & NMAD $m_i<22$ & Outlier Rate (\%) $m_i<26$ & NMAD $m_i<26$ \\ \hline
COS & 3.3                                & 0.027                  & 4.5                                & 0.029                  \\
COS+IRAC & 3.3                                & 0.027                  & 4.5                                & 0.029                  \\
XMM & 3.2                                & 0.028                  & 5.3                                 & 0.029                  \\
XMM+IRAC & 3.4                                & 0.030                  & 6.2                                 & 0.031                 
\end{tabular}
\end{table*}

\subsection{Photometric Redshifts}

The procedure we follow for obtaining photometric redshift estimates is the same as that used in \citet{Adams20}, with the only exception being the use of observer-frame NIR selection instead of optical selection. To summarise here, we make use of the minimising $\chi^2$ code {\sc LePhare} \citep{Arnouts1999,Ilbert2006} to fit templates of galaxies, active galactic nuclei (AGN) and stars to our SExtractor derived aperture fluxes. These templates are modified for dust attenuation according to the \citet{Calzetti2000} extinction law with E(B-V) = 0, 0.05, 0.1, 0.15, 0.2, 0.3, 0.6, 1.0, 1.5. The result is a Probability Density Function (PDF) for the redshift and a simple classification as a likely galaxy, star or QSO-like object. The template sets used are those from \citet{Ilbert2009} and are sourced from \citet{Polletta2007} and from \citet{Bruzual2003}. To conduct object classification, template spectra for AGN from \citet{Salvato2009} and stars from \citet{Hamuy1992,Hamuy1994,Bohlin1995,Pickles1998,Chabrier2000} were also fit. 
Photometric errors are set to a minimum of 5 per cent in flux during the template fitting process, this is to alleviate the consequences of using templates that probe the colour space discretely while the real galaxy population is continuous.

\subsubsection{Zero-point calibration with spectroscopic samples}

The two fields have been targeted by a large number of spectroscopic campaigns which can be used to calibrate our methods and examine photometric redshift accuracy. We make use of the spectroscopic catalogue compiled by the HSC team\footnote{The source of the spectroscopic catalogue compiled by the HSC team can be found at: \url{https://hsc-release.mtk.nao.ac.jp/doc/index.php/dr1_specz/}}. Included are spectra from the VIMOS VLT Deep Survey \citep[VVDS;][]{LeFevre2013}, z-COSMOS \citep{Lilly2009}, Sloan Digital Sky Survey \citep[SDSS-DR12;][]{Alam2015}, 3D-HST \citep{Skelton2014,Momcheva2016}, Primus \citep{Coil2011,Cool2013} and the Fiber-Multi Object Spectrograph \citep[FMOS;][]{Silverman2015}. From these we select only those with high quality flags (>95 per cent confidence) to ensure secure redshifts are being used. Together these provide a spectroscopic sample of 22,409 in COSMOS and 35,125 in XMM-LSS.

We use these spectroscopic redshifts to examine the accuracy of our photometric redshift estimates. 
In addition to this, we also make use of {\sc LePhare} functionality to make iterative adjustments to the zero-points of each photometric filter in order to optimise the results against the spectroscopic sample. Minor shifts in the zero-points can occur as a result of inaccurate filter transmission functions, through biases within the choice of SED templates and from the calibration of the images. The inclusion of a very large sample of spectroscopically confirmed objects from a variety of different surveys minimises the risk of introducing an additional bias through calibration on a non-representative sample. For each catalogue, we run {\sc LePhare} once on the spectroscopic sample to obtain the zero-point corrections, these offsets are applied and the entire field is then run. We show the results of this process in Table~\ref{tab:zeropoint} and the majority are small compared to the errors ($\ll 0.1$~mags).

\subsection{Photometric redshift accuracy}\label{sec:photoz-accuracy}

The quality of the photometric redshift catalogues can be described with two numerical values. The outlier rate, defined as percentage of objects satisfying abs($z_{spec}-z_{phot}$)/($z_{spec} + 1$) $> 0.15$, and the Normalised Mean Absolute Deviation \citet[NMAD;][]{hoaglin2000understanding}, defined as $1.48 \, \times\,$median[|$z_{spec}-z_{phot}$|/$(1+z_{spec})$]. These two values quantify a) the rate at which the photometric redshift method produces a redshift value that is in significant tension to the measured spectroscopic redshift and b) the spread around $z_{spec}-z_{phot}$ in a manner that is resistant to influence from the relatively small number of extreme outliers.
Comparing the two fields, COSMOS has greater depth and uniformity while XMM-LSS is shallower and wider, and has around 1~magnitude variability in its optical coverage . It is therefore expected that XMM-LSS would produce photometric redshifts of lower quality. The results of this comparison are displayed in Table~\ref{tab:stats}. For each field, we produce two sets of photometric redshift estimates, one with and one without the inclusion of the \textit{Spitzer} IRAC 3.6 and 4.5\,$\mu$m bands. The addition of the \textit{Spitzer} IRAC bands to COSMOS makes minimal difference in the quality of the photometric redshifts. However, redshifts are found to decrease in quality for the faintest objects in the XMM-LSS catalogue.

\subsection{Stellar mass determination}

With photometric redshifts and object classification determined for each source, we proceed to measure the stellar mass. This is performed by fixing the redshift to the best-fit value (template and redshift with minimum $\chi^2$) and rerunning LePhare using the total flux measurements, rather than aperture fluxes.  
For SExtractor we use fluxes from the \texttt{MAG\_AUTO} parameter and for ProFound this is \texttt{magt}. Compared to the aperture fluxes, the total fluxes are essential to making accurate measurements of the normalisation of the SED for resolved objects and hence the total luminosity of the galaxy and stellar mass. In the case that an object has a spectroscopic redshift from one of the surveys described in Section 3.2.1, this value is used instead of the photometric redshift.

For the SED fitting and hence, stellar mass determination, we use a selection of SED templates from \citet{Bruzual2003}. These include templates with constant star formation rate, exponentially decaying star formation rates with timescales $\tau = [0.1, 0.3, 1, 2, 3, 5, 10, 15, 30]$ Gyr, 58 ages for the stellar population (from 0.01 to 13.5 Gyr) and two metallicities ($Z_\odot$ and 20 per cent of $Z_\odot$).

\begin{comment}
\begin{table}
\caption{The primary statistics describing the photometric redshift quality of the catalogues used. The catalogues are split by field and the suffix `+Spitzer' indicates that \textit{Spitzer} IRAC 1/2 bands were included in the photometric redshift calculations.}
\label{tab:quality}
\begin{tabular}{l|ll}
        & Outlier Rate (\%) & NMAD  \\ \hline
COSMOS & 4.2               & 0.031 \\
COSMOS+Spitzer & 4.2               & 0.030 \\
XMM-LSS & 5.5               & 0.032 \\
XMM-LSS+Spitzer & 5.6               & 0.037
\end{tabular}
\end{table}
\end{comment}

\section{Measuring the GSMF}

We select objects for use in measuring the GSMF based on a number of criteria to maintain purity and completeness. 

\begin{enumerate}
    \item The source exists in both SExtractor and ProFound derived catalogues. The majority of sources that fail this are either artefacts on the edge of manual masking or are a consequence of the initial ProFound selection on a VISTA stack verses just the $K_s$-band. Such sources are also removed by the $K_s$ magnitude cuts detailed below.
    \item The source has a 2 arcsecond aperture magnitude following the condition $K_s < 24.5$ in COSMOS and $K_s < 23.2$ in XMM-LSS. This corresponds to a SNR cut of $8\sigma$ and is employed to minimise the potential for contamination while enabling $M^*$ to be well constrained up to $z\sim2$.
    \item The source has a best fit SED that is a galaxy or AGN with a redshift between 0.1 and 2.0 ($\chi^2_{\text{Gal/QSO}} < \chi^2_{\text{Star}}$). In the case a source has a spectroscopic redshift, that value is used in place of the photometric redshift.
    \item We apply an upper limit on the quality of the photometric redshift of $\chi^2 < 250$ (removing the worst 0.5 per cent of objects).
\end{enumerate}

This results in a sample of $\sim320,000$ galaxies in the combined COSMOS and XMM-LSS fields used in measuring the GSMF. We present the galaxy number counts in each redshift bin in Tables~\ref{tab:SE} and \ref{tab:PF}. The inclusion of \textit{Spitzer} bands leads to a significant reduction in the number of highly massive galaxies at redshifts $0.1 < z < 2$. This is the result of \textit{Spitzer} bands breaking degeneracies between stellar templates and certain combinations of galaxy template, leading to a number of these massive objects being reclassified as stars. The use of ProFound photometry over SExtractor leads to minimal difference to the general population of objects, cases will be discussed in Section \ref{sect:IRACimpact}.

\subsection{The 1/$V_{\rm max}$ method}

We first compute the GSMFs using the 1/Vmax methodology \citep{Schmidt1968,rowanrobinson1968}. 
We determined the $V_{\rm max}$ for each galaxy by redshifting the best-fitting template (from the 2 arcsecond aperture photometry) until the object no longer meets our $Ks$-band magnitude limit.

The GSMF is then determined using:

\begin{equation}\label{eqn:lf}
\Phi(M) d \log(M) = \frac{1}{\Delta M } \sum_i^N \frac{1}{V_{{\rm max},i}} ,
\end{equation}

\begin{equation}
\delta \Phi(M) = \frac{1}{\Delta M} \sqrt{\sum_i^N \left(\frac{1}{V_{{\rm max},i}}\right)^2} ,
\end{equation}
where $M$ is stellar mass,  $\Delta M$ is the bin width which we set to 0.25 dex and $V_{{\rm max},i}$ is the maximum volume for which galaxy $i$ could have been successfully detected.

\subsection{Stellar mass completeness}

Towards lower stellar masses, galaxies become intrinsically less luminous. This ultimately leads to a regime where the detection limits of the data are reached and galaxy number counts begin to fall as they are lost to noise. 
As the science goals of this study focus on the massive end of the GSMF, we adopt a conservative approach while still probing a significant mass range. In our model fitting procedures we elect to only use bins of redshift and mass where over 95 per cent of the galaxy sample are brighter than the $8\sigma$ 2~arcsecond aperture detection limit of the respective field. Due to the shallower coverage in XMM-LSS, the criteria for completeness is approximately 0.5-0.75~dex higher in stellar mass than in COSMOS.

We apply a simple correction to the survey areas based on the fraction of the field that is occupied by other sources or masked by foreground stars. For COSMOS this is 15 per cent and XMM it is 7 per cent. This corrects the GSMF for the probability that sources are highly blended ($>50$ per cent overlap) with other sources or significantly effected by bright foreground stars.

\subsection{Cosmic Variance}

\begin{figure}%
    \centering
    \includegraphics[width=\columnwidth]{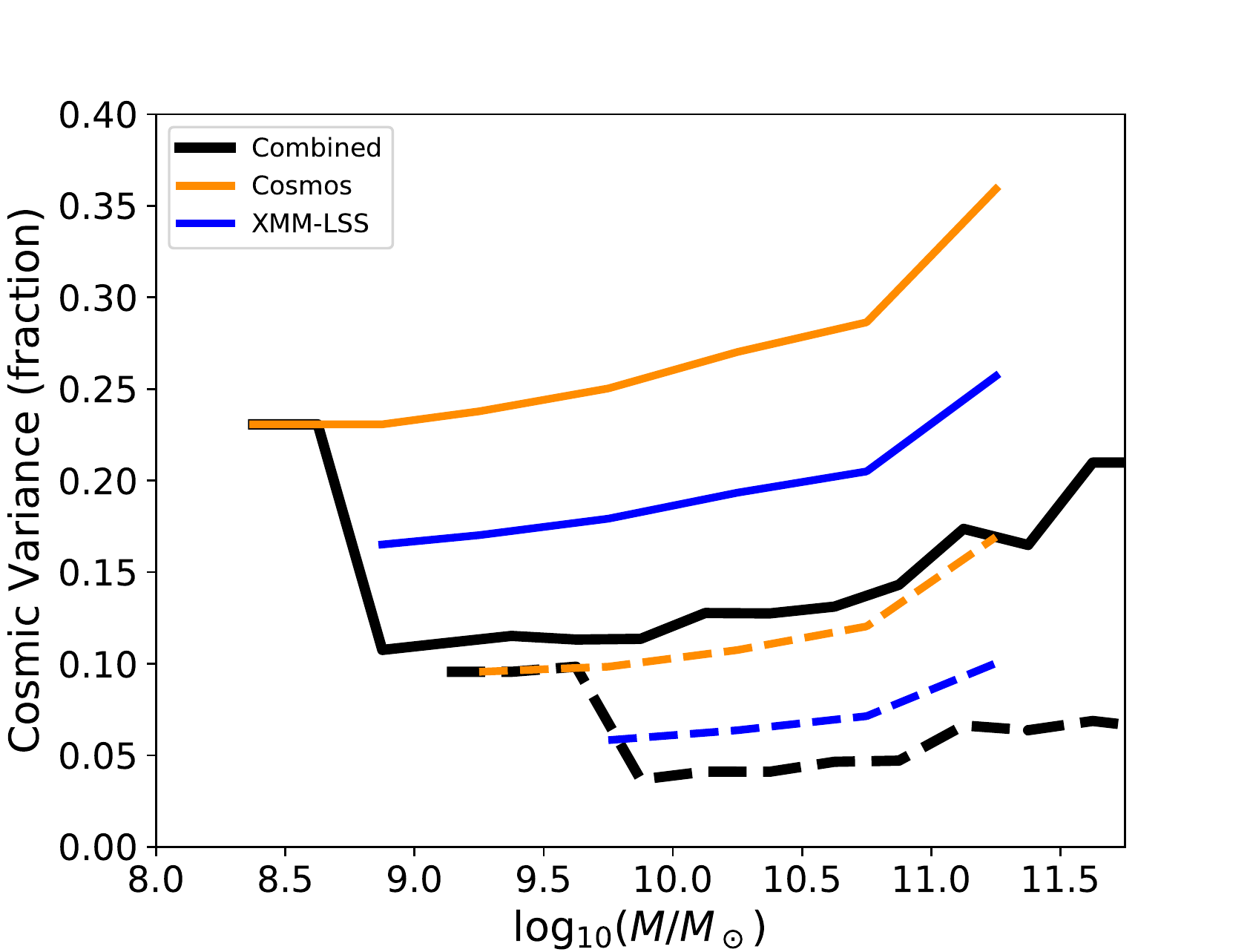}
    \caption{The estimated cosmic variance for two example redshift bins. The solid lines show $0.2 < z < 0.3$, the dashed lines show $0.75 < z < 1.0$. Here we display the cosmic variance for each field if treated independently and the result of combining the fields using the cosmic variance calculator from \citet{Moster11} and our measured number counts. COSMOS is shown in gold, XMM-LSS in blue and the combination of the two fields in black. Where XMM-LSS becomes incomplete the cosmic variance value for the combined case is just the COSMOS cosmic variance.}
    \label{fig:CV}
\end{figure}

Our measurements of the GSMF are based on data that only probe a limited volume of the Universe. As a result, they are susceptible to biases that are a consequence of large-scale fluctuations in density in the galaxy distribution. This is commonly referred to as cosmic variance ($\sigma_{cv}^2$). As we are measuring the GSMF across a wide range of mass and redshift, there is no single quantitative value that can be used to describe this effect. In order to model the effects of cosmic variance on our measurements, we use the treatment from \citet{Moster11}, which provides an estimate of the cosmic variance as a function of both stellar mass and redshift. Our dataset consists of two fields with differing area and dimensions, thus allowing us to mitigate some of the effects of cosmic variance. Where both the XMM-LSS and COSMOS fields are used in measuring the GSMF we calculate the cosmic variance for each field independently ($\sigma_{cv,i}^2$) and combine the values together with a co-moving volume weighting \citep[Equation 7 in][]{Moster11}. The output is a percentage error on the GSMF, and so to combine the fields this value is converted back into variance ($\sigma_{cv}^2$) using our galaxy number counts. When data from the XMM-LSS field drops out of consideration due to its shallower depth, cosmic variance is determined from the area of the COSMOS field alone. Fig.~\ref{fig:CV} shows the results of our cosmic variance calculations for two redshift bins ($0.2 < z < 0.3$ and $0.75 < z < 1.0$), highlighting how the increased area from including the XMM-LSS field, coupled with combining two widely separated fields, results in a factor of $\sim 2$ decrease in the cosmic variance uncertainty. 

\begin{figure}%
    \centering
    \includegraphics[width=\columnwidth]{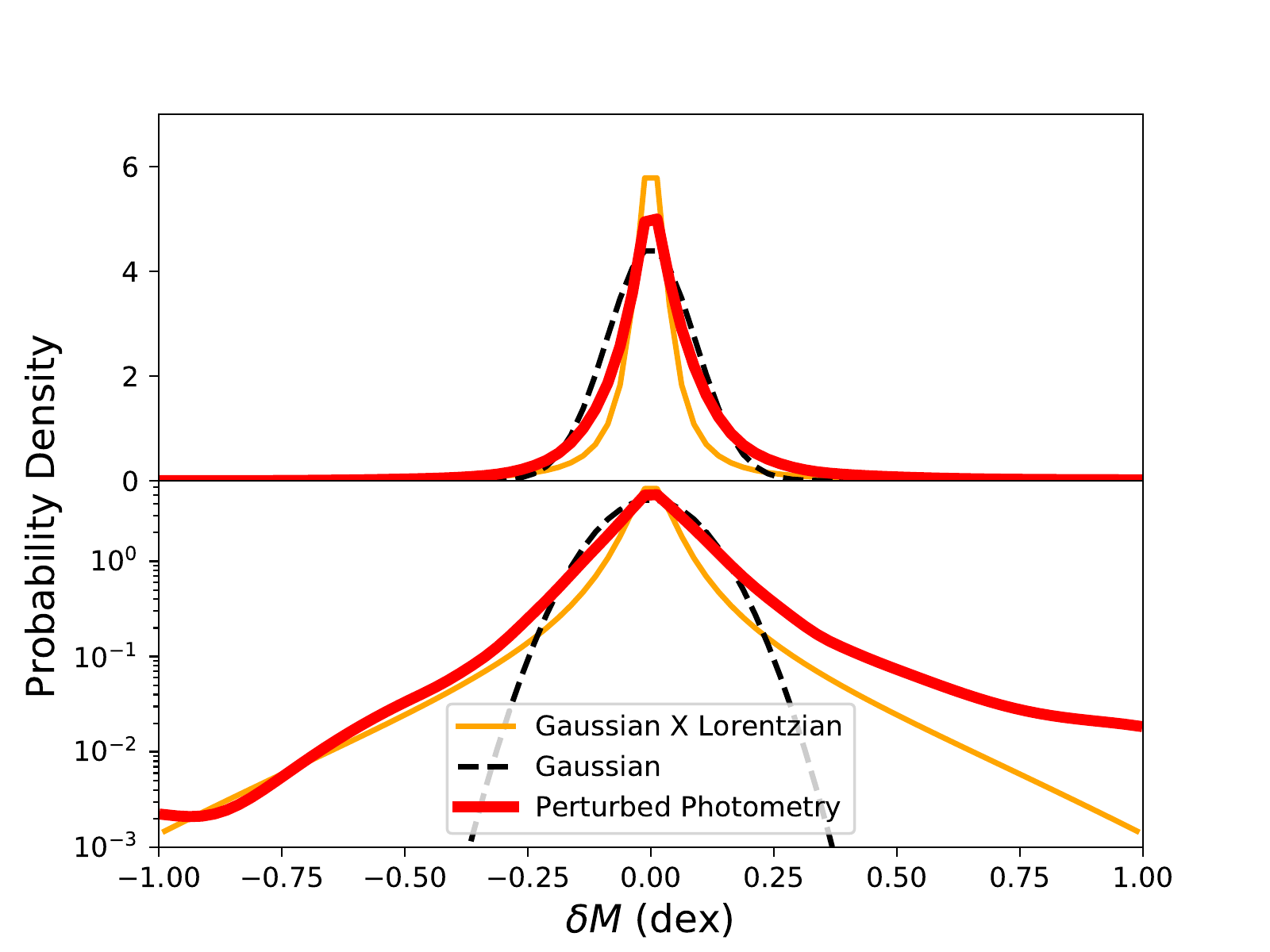}
    \caption{The shapes of the three Eddington bias corrections implemented as a part of our study. We show the probability of a certain mass scatter ($\delta M = M_{\text{original}} - M_{\text{perturbed}}$) as a result of photometric errors in the SED fitting process.  The lower plot shows identical data to the upper plot, however we have logged the $y$-axis to reveal the low-amplitude wings of the distribution.  The non-parametric distribution derived directly from our data is shown as the thick red line.  The black dashed line is the best fit Gaussian to this data and the orange line is for a Gaussian multiplied by a Lorentzian.}
    \label{fig:Edd}
\end{figure}

\begin{figure*}%
    \centering
    \includegraphics[width=2\columnwidth]{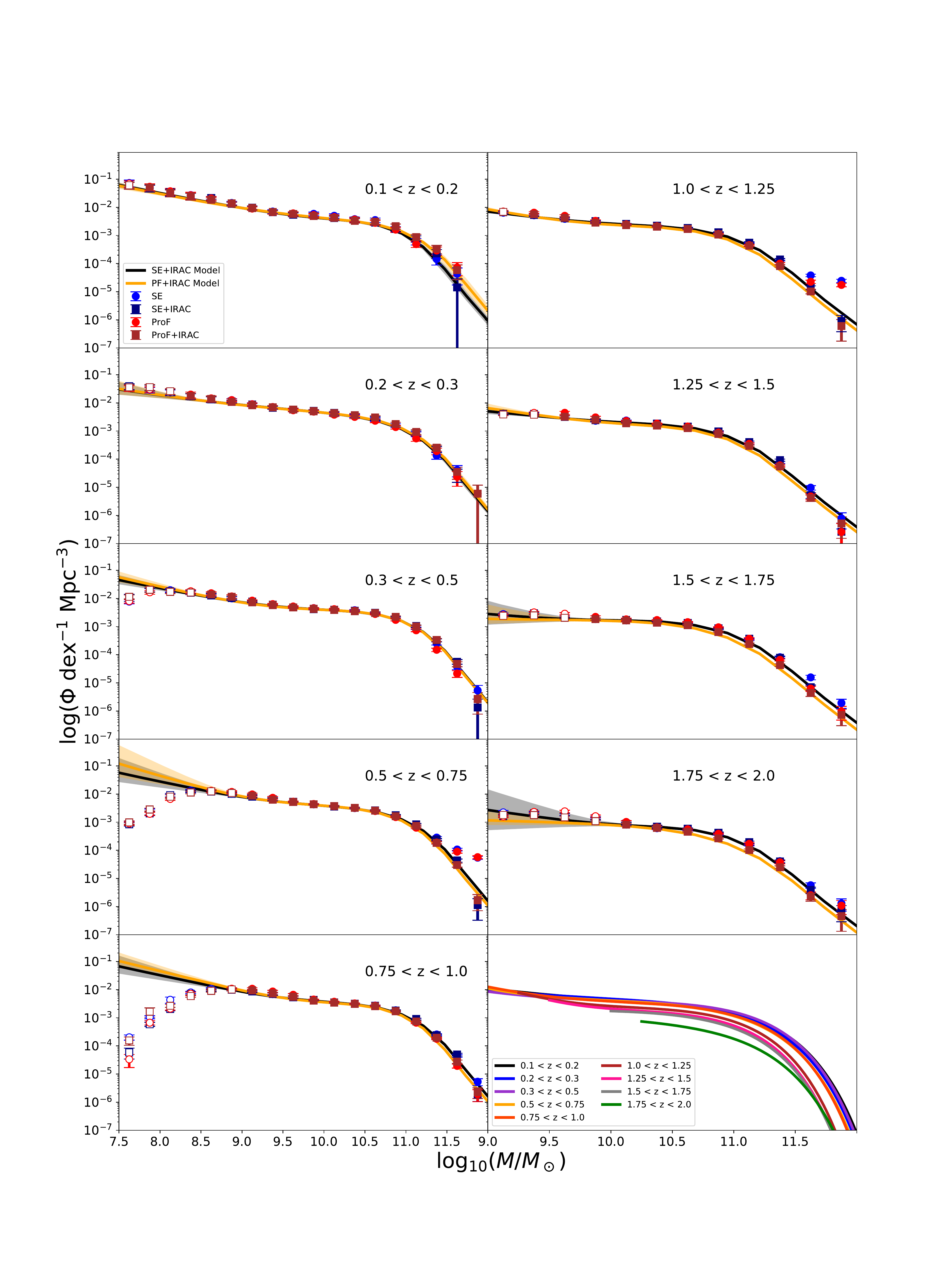}
    \vspace{-1.8cm}
    \caption{The GSMF from our analysis of the COSMOS and XMM-LSS data. In blue and red are the results from SExtractor and ProFound, while navy and brown data points denote the points measured after including \textit{Spitzer}/IRAC photometry. Unfilled data points indicate those with less than 95 per cent completeness and are not used in any fitting procedure. Displayed data points are based on raw observations and do not correct for Eddington bias (see Tables \ref{tab:SE} and \ref{tab:PF}). The black line shows the median of the MCMC results for the SExtractor+IRAC data when fit with a double Schechter function convolved with our Eddington correction. The grey shaded region shows the region contained within $1\sigma$ of the model fit and is based on 10,000 random samples of the final posterior. The gold line and shaded region follows the same process with the ProFound+IRAC data. The high redshift bins in the right column have the x-axis truncated to higher masses to focus on the complete regime. In the final panel the Eddington corrected ProFound models are shown simultaneously and cut where data is incomplete.}
    \label{fig:MF}
\end{figure*}

\subsection{Eddington bias}\label{sect:eddmethod}

The steep drop in the GSMF beyond the knee 
can lead to a bias in the derived number densities of the most massive galaxies due to Eddington bias \citep{Eddington1913}.
All galaxies in the sample have an uncertainty in the derived stellar mass, however as low-mass galaxies are significantly more numerous this leads to more galaxies scattering to higher stellar mass than the number that scatter to lower masses. 
To account for this effect and hence, determine an estimate of the intrinsic GSMF, we require an estimate of the uncertainty in the stellar masses derived for our sample.
With this distribution we can then deconvolve (or in reality, fit a convolved double Schechter form) to our observations to determine the intrinsic GSMF.
To measure the uncertainty in the stellar masses derived in this study, we repeat both the photometric redshift measurement and the SED fitting process after perturbing the photometry of our sources according to the photometric errors in each band. This process is repeated multiple times to produce the distributions shown in Fig.~\ref{fig:Edd}.
Based on this, we examine three possible methodologies to uncover the intrinsic stellar mass function from this observed distribution in our analysis described in Section~\ref{sec:measuredGSMF}.

\subsection{The measured galaxy stellar mass functions}\label{sec:measuredGSMF}

The GSMFs, as measured from the samples produced from our four catalogues, are presented in Fig.~\ref{fig:MF}. They probe stellar masses from $7.75 < \log_{10}(M_\odot) < 11.75$ and are split into nine redshift bins between $0.1 < z < 2.0$. 

To each GSMF we fit a double Schechter function \citep[][Eqn.~\ref{eqn:schecht}]{Schechter1976} using a Markov-Chain Monte Carlo (MCMC) implemented in {\tt emcee} \citep{emcee}. In this redshift range, past studies have found the double Schechter functional form better describes the GSMF due to the underlying bimodality in the galaxy population \citep[e.g. ][]{Strateva2001,Driver2006,Baldry2012,Ilbert2013,Davidzon17} and we clearly see an upturn at the low mass end of our GSMFs.

\begin{equation}\label{eqn:schecht}
    \Phi(M)\delta M = \left[ \Phi_1 \left( \frac{M}{M^*} \right)^{\alpha_1} +  \Phi_2 \left( \frac{M}{M^*} \right)^{\alpha_2}  \right] \exp\left(-\frac{M}{M^*}\right)\frac{\delta M}{M^*}.
\end{equation}

A series of priors are applied to prevent parameters from flipping due to the symmetry of the double Schechter functional form shown in Equation~\ref{eqn:schecht}. The normalisation ($\Phi_1$ \& $\Phi_2$) follows the condition $\Phi_1 > \Phi_2$, the low-mass slopes ($\alpha_1$ \& $\alpha_2$) are limited to being between [-1.8,1.5] and [-3.0,-0.9] respectively for the two components. To compare against past studies, we make the same assumption that each Schechter component has a single, shared value of $M^*$ in the range $10< \log_{10}(M^*/{\rm M}_\odot) < 12$. We only fit to data points where the bins in stellar mass are greater than 95 percent complete. 
The MCMC is set up with 500 walkers that burn in for 100,000 steps before conducting a further 20,000 steps for use in mapping the posterior distributions. Each walker is distributed uniformly in the parameter space and limited by the above priors. 
We perform the fitting procedure four times, once on the observed GSMF and three times with the double Schechter function modified with the convolution with one of three Eddington bias methods that we describe below.

First, we modify the fit to be a convolution of the double Schechter function with a Gaussian distribution, where the standard deviation ($\sigma_M$) of the distribution is calculated by fitting a Gaussian to the measured scatter in masses shown in Fig.~\ref{fig:Edd}. 
This is a commonly used method in studies of the GSMF \citep[e.g.][]{Wright18}.
In the second method we extend this model by multiplying the Gaussian with a Lorentzian in the same manner as described in \citet{Ilbert2013,Davidzon17}. This adds extended wings to the function, which more adequately reproduces the distribution, however it continues with the assumption that the mass scatter is symmetric. Any asymmetry is important to account for, as it means that there is a greater probability of scattering to lower masses than towards higher masses (due to the photometric redshift uncertainty), and this would impact on the measured GSMF. Therefore, in the third case, we do not fit any parametric distribution to the mass scatter, instead we use the smoothed histogram of the scatter convolved directly with the double Schechter function. This should  improve upon the use of the analytic forms because it captures the observed asymmetry in the mass scatter ($\delta M$).

Visual inspection of the distribution in mass scatter derived in Section~\ref{sect:eddmethod} reveals there to be minimal dependence on redshift. The two lowest redshift bins are slightly broader as a consequence of photometric redshift uncertainty, but these bins contain a much higher fraction of galaxies with spectroscopic redshifts (30 per cent). As a result, the real scatter within these bins is likely much smaller. Using the method which convolves a Gaussian distribution with the measured GSMF, we find values with $0.08 < \sigma_M < 0.10$ across all redshifts.  For the second method, which uses a Gaussian distribution multiplied with a Lorentzian, we find $0.43 < \sigma_M < 0.58$.  We note that the $\sigma_M$ values for the two methodologies are not directly comparable due to the different functional forms. Since minimal evolution was found, we remove the redshift dependence on the Lorentzian component that was introduced in \citet{Ilbert2013} to minimise total fit parameters. The resultant formula for the extended wings is thus $L(x) = \frac{0.08}{2\pi}\frac{1}{\frac{0.08}{2\pi}^2+x^2}$, which is equivalent to fixing the redshift to 1.0 in the original formula from \citet{Ilbert2013}.  The range of $0.43 < \sigma_M < 0.58$ for the Gaussian component is in agreement with the findings of \citet{Ilbert2013} and slightly larger than the value found by \citet{Davidzon17}. In addition, \citet{Grazain2015} and \citet{Davidzon17} both find evidence for redshift and stellar mass dependence on the measured mass scatter when approaching the completeness limited regime. The probable explanation for the lack of such a dependence in our data is the conservative SNR cuts that have been implemented.

Following previous studies, we fix the $\sigma_M$ values in our final fits to 0.09 in the Gaussian case and 0.5 for the Gaussian $\times$ Lorentzian case. The shapes of these distributions are presented in Fig.~\ref{fig:Edd}. We discuss the impacts of this correction on the measured GSMF in Section~\ref{sect:edd}.

Our preferred method for correcting for Eddington bias is to use the histogram presented in Fig.~\ref{fig:Edd} directly in the convolution. This method directly uses the results of the perturbed catalogues and captures the subsequent asymmetry found in the distribution. Such an asymmetry has previously been described in recent studies exploring the Eddington bias \citep{Grazain2015}.
The best-fitting double Schechter function fit parameters using this Eddington bias correction are presented in Table~\ref{tab:Results}. Corner plots showing the posterior probability distribution for the parameters in this model will be provided in an online resource. 
For completeness, we also present our results without the application of the Eddington bias correction in Appendix \ref{alternatefits} alongside the results obtained using the various parametric fits to the mass scatter.

Our double Schechter model is only fit up to stellar masses of $10^{11.75}M_\odot$. While there are a small number of objects with stellar masses greater than this limit in our sample, these are increasingly likely to be subject to forms of contamination such as AGN activity, source blending, misclassification of stars or artefacts. In the high redshift regime we are also unable to constrain the low-mass Schechter component. If we instead fit with a single Schechter function for $z>1.25$, we find the shift in the fit parameters to be minimal ($<1\sigma$) compared to results obtained with a double Schechter function. For consistency we proceed with the double Schechter functional form for all redshift bins.

Visual inspection of the measured GSMFs reveal a change in the shape of the massive-end between the redshift bins of $0.75<z<1.0$ and $1.0<z<1.25$ (see the last panel of Fig.\ref{fig:MF}). This evolution is present in the results obtained using both source extraction methods (SExtractor/ProFound) and both sets of photometric redshifts (including/excluding \emph{Spitzer} data). Inspection of the redshift distributions reveal no significant features, such as peaks or troughs, within these redshift bins. The total shift in the high-mass component amounts to around 2 sigma in $M^*$ and the normalisation $\Phi_1$ between these two redshift bins. So, a combination of statistical errors/cosmic variance or an unknown systematic could be the driver of such a change.

\begin{table*}
\caption{The best-fitting double Schechter function parameters derived from our observed GSMF when corrected for Eddington bias through the direct implementation of the mass scatter over fitting an analytic form. Priors are set to ensure that $\alpha_1$ \& $\log_{10}(\Phi_1)$ refer to the high mass component and $\alpha_2$ \& $\log_{10}(\Phi_2)$ refer to the low mass component.}
\label{tab:Results}
\begin{tabular}{c|ccccc}
\hline
Redshift & $\log_{10}(M^*/M_\odot)$& $\alpha_1$ & $\log_{10}(\Phi_1$) & $\alpha_2$ & $\log_{10}(\Phi_2)$  \\ SExtractor+IRAC & & & $[\textrm{mag}^{-1} \textrm{Mpc}^{-3}]$ & & $[\textrm{mag}^{-1} \textrm{Mpc}^{-3}]$   \\
\hline
0.1-0.2  &  $10.71^{+0.11}_{-0.11}$      & $-0.47^{+0.44}_{-0.36}$    & $-2.61^{+0.10}_{-0.12}$   & $-1.60^{+0.09}_{-0.12}$    &  $-3.34^{+0.23}_{-0.36}$  \\[4pt]
0.2-0.3  &  $10.89^{+0.08}_{-0.13}$      & $-1.06^{+0.59}_{-0.17}$    & $-2.76^{+0.12}_{-0.10}$   & $-1.57^{+0.24}_{-0.65}$    & $-4.12^{+1.03}_{-3.28}$   \\[4pt]
0.3-0.5  &  $10.83^{+0.05}_{-0.05}$      & $-0.64^{+0.22}_{-0.17}$    & $-2.60^{+0.05}_{-0.05}$   & $-1.61^{+0.14}_{-0.18}$    & $-3.62^{+0.30}_{-0.44}$ \\[4pt]
0.5-0.75  &  $10.86^{+0.05}_{-0.05}$      & $-0.80^{+0.30}_{-0.18}$    & $-2.71^{+0.05}_{-0.06}$   & $-1.69^{+0.27}_{-0.47}$    & $-3.78^{+0.56}_{-1.06}$   \\[4pt]
0.75-1.0  &  $10.83^{+0.05}_{-0.05}$      & $-0.67^{+0.28}_{-0.20}$    & $-2.68^{+0.04}_{-0.05}$   & $-1.67^{+0.19}_{-0.32}$    & $-3.63^{+0.38}_{-0.68}$   \\[4pt]
1.0-1.25  &  $10.67^{+0.05}_{-0.05}$      & $-0.19^{+0.35}_{-0.30}$    & $-2.75^{+0.05}_{-0.07}$   & $-1.58^{+0.18}_{-0.28}$    &  $-3.39^{+0.25}_{-0.44}$ \\[4pt]
1.25-1.5  &  $10.60^{+0.05}_{-0.05}$      & $0.00^{+0.37}_{-0.42}$    & $-2.87^{+0.09}_{-0.10}$   & $-1.44^{+0.18}_{-0.47}$    &  $-3.29^{+0.20}_{-0.60}$  \\[4pt]
1.5-1.75  &  $10.66^{+0.05}_{-0.06}$      & $-0.45^{+0.40}_{-0.23}$    & $-2.84^{+0.04}_{-0.09}$   & $-1.79^{+0.59}_{-0.78}$    &  $-4.39^{+1.01}_{-2.39}$  \\[4pt]
1.75-2.0  &  $10.66^{+0.07}_{-0.08}$      & $-0.25^{+0.54}_{-0.42}$    & $-3.19^{+0.04}_{-0.10}$   & $-1.94^{+0.60}_{-0.70}$    &  $-4.36^{+0.69}_{-2.88}$  \\[4pt] \hline
ProFound+IRAC & & & & &  \\
\hline
0.1-0.2  &  $10.88^{+0.16}_{-0.15}$      & $-0.74^{+0.50}_{-0.40}$    & $-2.71^{+0.15}_{-0.20}$   & $-1.62^{+0.12}_{-0.23}$    &  $-3.54^{+0.35}_{-0.91}$  \\[4pt]
0.2-0.3  &  $10.90^{+0.09}_{-0.13}$      & $-1.04^{+0.56}_{-0.18}$    & $-2.75^{+0.12}_{-0.11}$   & $-1.60^{+0.26}_{-0.64}$    &  $-4.13^{+1.02}_{-3.05}$  \\[4pt]
0.3-0.5  &  $10.83^{+0.05}_{-0.05}$      & $-0.67^{+0.21}_{-0.16}$    & $-2.58^{+0.04}_{-0.05}$   & $-1.70^{+0.17}_{-0.22}$    &  $-3.79^{+0.38}_{-0.52}$   \\[4pt]
0.5-0.75  &  $10.81^{+0.04}_{-0.05}$      & $-0.86^{+0.25}_{-0.13}$    & $-2.67^{+0.04}_{-0.05}$   & $-1.94^{+0.39}_{-0.54}$    &  $-4.21^{+0.78}_{-1.09}$  \\[4pt]
0.75-1.0  &  $10.75^{+0.04}_{-0.04}$      & $-0.54^{+0.22}_{-0.18}$    & $-2.62^{+0.03}_{-0.04}$   & $-1.76^{+0.17}_{-0.24}$    &  $-3.64^{+0.30}_{-0.45}$ \\[4pt]
1.0-1.25  &  $10.63^{+0.05}_{-0.05}$      & $-0.29^{+0.31}_{-0.24}$    & $-2.74^{+0.03}_{-0.05}$   & $-1.81^{+0.22}_{-0.31}$    &  $-3.64^{+0.32}_{-0.50}$  \\[4pt]
1.25-1.5  &  $10.57^{+0.05}_{-0.05}$      & $-0.08^{+0.39}_{-0.34}$    & $-2.89^{+0.06}_{-0.10}$   & $-1.65^{+0.24}_{-0.43}$    &  $-3.46^{+0.26}_{-0.55}$   \\[4pt]
1.5-1.75  &  $10.66^{+0.04}_{-0.06}$      & $-0.70^{+0.32}_{-0.13}$    & $-2.91^{+0.04}_{-0.05}$   & $-1.86^{+0.62}_{-0.75}$    &  $-5.05^{+1.35}_{-3.19}$   \\[4pt]
1.75-2.0  &  $10.76^{+0.05}_{-0.05}$      & $-0.98^{+0.19}_{-0.13}$    & $-3.38^{+0.06}_{-0.07}$   & $-1.86^{+0.64}_{-0.74}$    &  $-6.64^{+2.07}_{-2.29}$ \\[4pt] \hline
\end{tabular}
\end{table*}

\section{Results}

\subsection{Changes in the GSMF with varying methodology}

Within each redshift bin, we measure a total of four stellar mass functions. First we compare the results with and without the inclusion of~\emph{Spitzer}/IRAC [3.6] and [4.5] data, and second we compare the GSMF derived from SExtractor based photometry in comparison to that derived with Profound.
While the low mass component of the GSMF is consistent between these different methods, we find some differences in the results at stellar masses greater than $M^*$.

\subsubsection{The impact of including \textit{Spitzer}/IRAC photometry on the measured GSMF}\label{sect:IRACimpact}

The most immediately apparent feature is the offset between the GSMFs that include or exclude \textit{Spitzer}/IRAC data in the redshift bins of $0.5 < z < 0.75$ and $1.0 < z < 1.25$. This is due to two effects. Firstly, many high-mass objects that lie in the $0.5 < z < 0.75$ bin have a different redshift solution ($z \sim 0.1$) when \textit{Spitzer}/IRAC is included. They consequently have smaller masses in this lower redshift bin and their influence is negated by the high number densities within this mass-redshift range. The cause of the different redshift solutions is the uncertainty of the SED slope redder than the $K_s$ band, with a red slope favouring the higher redshift solutions and a blue slope favouring low redshift solutions. 
Secondly, there is a likely contamination from stars in the $0.5 < z < 0.75$ and $1.0 < z < 1.25$ bins. Many high mass objects have smooth red optical slopes that turn over around the $H$ or $K_s$ bands. With a limited wavelength range, some black body-like spectra and red galaxies are indistinguishable. Introducing the \textit{Spitzer}/IRAC bands vastly increases the $\chi^2$ of the galaxy models and reduces the $\chi^2$ of the stellar models for a significant number of these high mass objects. Consequently these objects no longer meet our selection criteria (either the $\chi^2$ increases above 250 or the classification switches to a star) when using the values associated with the inclusion of the mid-infrared bands. Both of these cases are examples of degeneracies between template sets as a result of the use of a limited number of broadband filters. 
Thus we find that the inclusion of the [3.6] and [4.5] bands makes a significant difference to the derived number density of the most massive galaxies in our data.
Therefore, we focus on the GSMFs measured with the inclusion of the \textit{Spitzer}/IRAC when discussing redshift evolution and comparisons to simulations.

\subsubsection{The impact of varying the choice of source extraction software}\label{sect:ProFimpact}

While the global mass distributions between SExtractor and ProFound based catalogues are broadly very similar, there are individual cases where mass estimates can vary widely between objects. 
Issues can typically be put down to artefacts affecting the photometry, proximity to bright sources and disagreement between the two source extraction measurements in the {\em Spitzer} bands. Instances of significant differences in mass estimations tend to occur in high redshift and/or low mass objects that fall within our incomplete regime and so do not affect our final results.

At low redshifts, galaxies become increasingly resolved and so any systematic variation between the SExtractor and ProFound photometry would be expected to be more apparent. In our measurements we find any such variation between the two to be very small, with the low-mass components ($M<M^*$) between $0.1 < z < 1.5$ being highly consistent between the two source extraction methods. We find that at the lowest redshift ($0.1<z<0.2$) the ProFound based GSMF produces more galaxies of very high mass compared to the SExtractor derived measurements. This is then reversed at higher redshifts $z>0.5$, where the ProFound GSMF produces fewer galaxies of very high mass. However, these differences are of relatively low significance (of order $1\sigma$) and demonstrate that over the redshift and mass ranges probed in this study, the choice of source extraction software makes no significant difference to GSMF results.

\subsection{The intrinsic GSMF corrected for Eddington bias}\label{sect:edd}

\begin{figure}%
    \centering
    \includegraphics[width=\columnwidth]{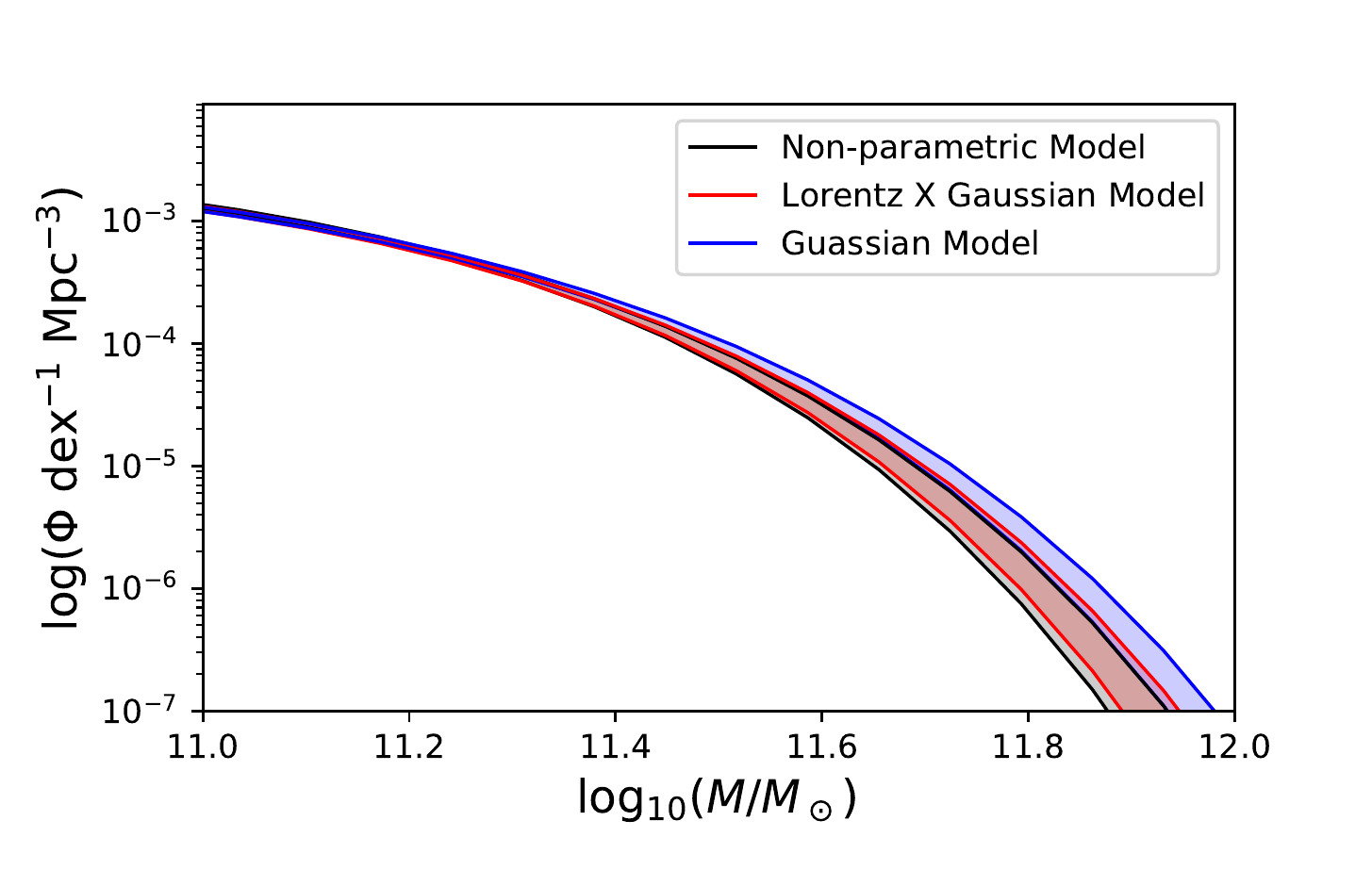}
    \caption{The impact that differing models of the Eddington Bias have on the measurement of the intrinsic GSMF at high masses. Shown is the intrinsic GSMF at $0.5<z<0.75$ as measured with SExtractor photometry when recovered using the three methods applied in this study. In blue we show the results of using a simple Gaussian to model mass errors, in red we expand the model to include Lorentzian wings and in black we show the results when using a non-parametric approach. Shaded regions indicate the 1 sigma uncertainty and are derived from 10,000 random samples of the posterior probability distribution. The edges of the shaded regions are made more bold to assist in readability.}
    \label{fig:EddImpact}
\end{figure}

To recover the intrinsic GSMF from our observations, which is affected by the uncertainties in the estimate of stellar mass, we trial three different methods described in Section~\ref{sect:eddmethod}. In the first method, we assume that the scatter induced by uncertainties on stellar mass are described by a Gaussian distribution of $\sigma_M = 0.09$. We find that this imposes minimal changes to the shape of the high-mass end of the MF, corresponding to a shift in $M^*$ of order 0.05~dex lower when compared to the observed GSMF. With the second method, where we convolve a Gaussian $\times$ Lorentzian combination with $\sigma_M = 0.5$, we find the shift between the observed and intrinsic parameters to be more significant with $M^*$ shifting to lower masses by $\sim 0.1$~dex compared to the observed GSMF. For completeness, we display the MCMC results for these two methods in Tables~\ref{tab:LogN} and \ref{tab:LxG} respectively.
Each of these methods make a fundamental assumption in that the scatter in the derivation of the stellar mass is symmetric in logarithmic stellar mass space. We find this assumption to work best when redshifts are confident e.g. if we conduct the photometry scatter procedure on just the mass calculations and assume the photometric redshift is correct or the objects have spectroscopic redshifts. However, when the uncertainty on the photometric redshifts is included, the measured distribution is found to be broader and more asymmetric. This results in the Gaussian $\times$ Lorentizan method underestimating the amount of scattering in certain regimes (small up-scatter and most of the down-scattering), even with the extended wings of the function. It is for this reason that we elect to instead use the measured distribution of mass scatter to convolve with the intrinsic double Schechter function (see Fig.~\ref{fig:EddImpact} for an example of the impact).

The GSMF, when corrected with this distribution in mass scatter, undergoes a similar shift in the Schechter parameters to that of the Gaussian $\times$ Lorentzian. The shift in $M^*$ is 0.12~dex compared to the observed GSMF and all the fit parameters lie within $1\sigma$ of the results found with the Gaussian $\times$ Lorentzian method. While the broad wings of the distribution of mass scatter are very small in probability beyond shifts of 0.3~dex (see Fig.~\ref{fig:Edd}), the nature of the GSMF spanning many orders of magnitude in number density around the knee requires these wings to be modelled in order to capture the impact of a small number of objects scattering into the less populated, high-mass bins. This highlights that the intrinsic GSMF is very sensitive to the strength of the Eddington bias correction and could be a key source of inconsistency between results of observational studies.  The best-fit parameters for the double Schechter function when using our preferred non-parametric Eddington correction are shown in Table~\ref{tab:Results}.

The use of high completeness spectroscopic surveys would reduce the uncertainty on stellar mass due to photometric redshift uncertainties \citep[][]{Pozzetti2010,Moustakas2013,Leauthaud2016}. While such studies have been limited to the brighter and more massive objects (e.g. $\log_{10}(M/M_\odot) > 10.5$), it is these objects that are most at risk at having their number counts inflated by Eddington Bias. With new surveys and instruments using multi-object spectrographs coming online in the coming years (e.g. DEVILS and WAVES), such studies will soon be capable of measuring the GSMF to lower masses and higher redshifts \citep{Davies18,WAVES}

\subsection{Comparisons to previous studies}

\begin{figure*}%
    \centering
    \includegraphics[width=2\columnwidth]{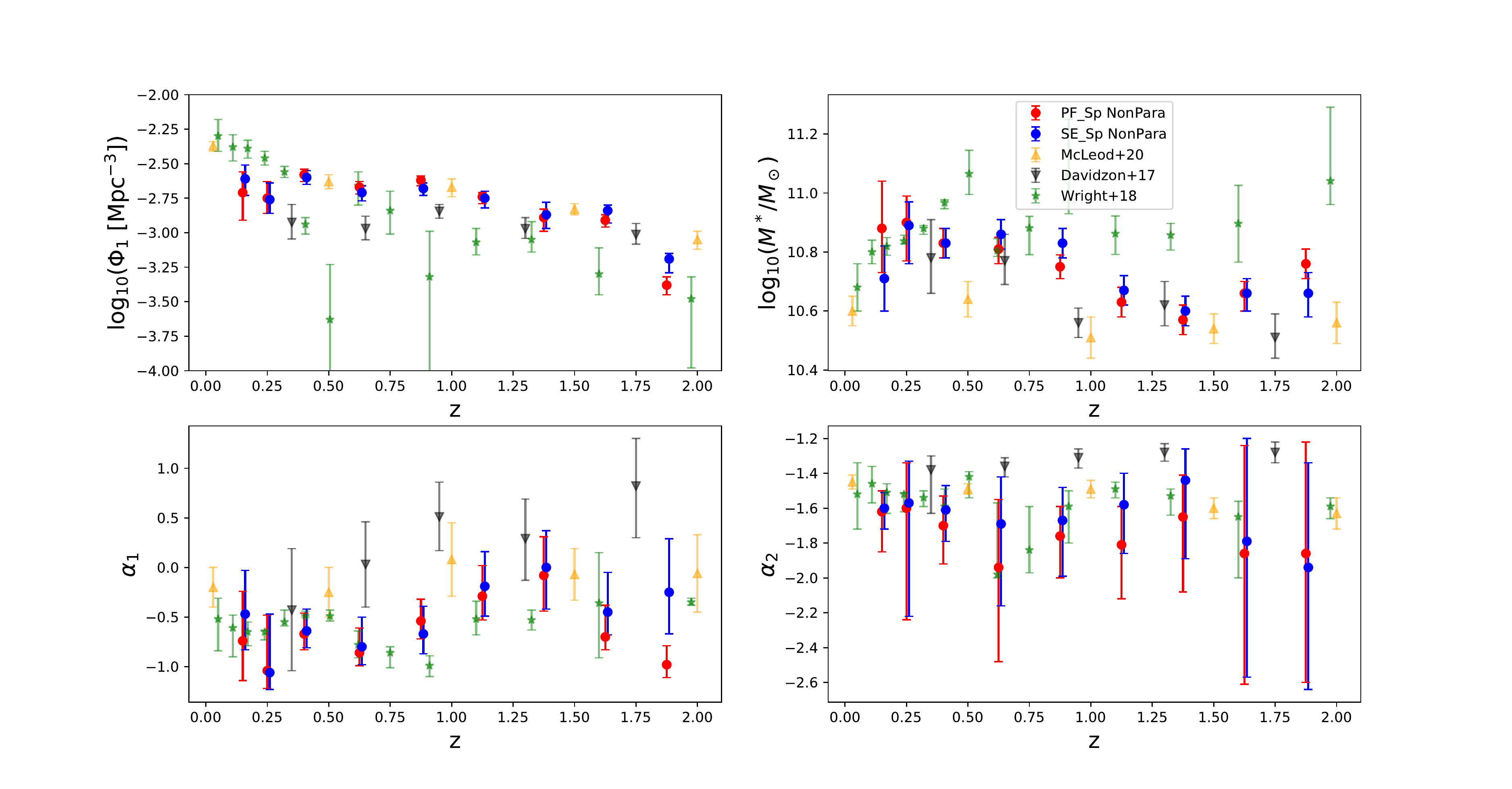}
    \caption{The time evolution of the double Schechter parameters as measured with photometry derived by SExtractor (blue) and ProFound (red) and including the use of \textit{Spitzer}/IRAC photometry in both redshift and mass calculations. These data points are for the fits that utilise the non-parametric correction for Eddington bias. Since the parameter $\Phi_2$ is largely unconstrained for a significant portion of the redshift range, we do not show those results in this figure. Alongside our results we display the results of \citet{Davidzon17,Wright18,McLeod2020} in black, gold and green respectively. The lowest redshift data point from \citet{McLeod2020} is a modified value from \citet{Baldry2012}}
    \label{fig:Mstar}
\end{figure*}

To put this study into greater context we compare against a number of past studies.
The studies selected for these comparisons are \citet{Davidzon17} which utilised only the COSMOS field in their study with the \citet{Laigle16} catalogue, \citet{Wright18} which uses a combination of the GAMA \citep{Driver2011}, COSMOS \citep{Davies2015, Andrews2017} and 3D-HST surveys \citep{Brammer2012,Skelton2014,Momcheva2016} and the recent study conducted by \citet{McLeod2020} that uses components of the COSMOS \& XMM-LSS fields. The results of the the double Schechter fits conducted in these studies are presented alongside our own in Fig.~\ref{fig:Mstar} and all were calculated with the same cosmological model as used in this study.

Examining the Schechter parameters individually, the strongest agreement is that of the high mass normalisation ($\Phi_1$), where there is very close agreement between our results and those from \citet{McLeod2020}. In contrast, the slope of the high mass component $\alpha_1$ is found to agree more with the results from \citet{Wright18}, however this is subject to degeneracy based effects from our poorer capability to constrain the low-mass slope $\alpha_2$. Inspection of the corner plots show that the more negative values of $\alpha_1$ are driven by the steeper slopes found in our broad uncertainties of the low-mass slope (see accompanying online resources).

The Eddington bias corrections implemented in previous work all use different functional forms. \citet{Wright18} and \citet{McLeod2020} use corrections that are independent of mass and redshift as in our study, whereas \citet{Davidzon17} implement a minor redshift dependence in the Lorentzian terms. Furthermore, \citet{Wright18} use a Gaussian of width $\sigma_M = 0.1$~dex while \citet{Ilbert2013,Davidzon17} use the Gaussian $\times$ Lorentzian distribution of widths 0.5~dex and 0.35~dex respectively. \citet{McLeod2020} uses a log-normal distribution of width $\sigma_M=0.15$~dex to correct for the Eddington bias, leading to larger changes in their fit parameters, with $M^*$ changing by up to 0.2~dex compared to the raw observations. Other examples of recent studies include \citet{Thorne2020} and \citet{Leja2020}, who fold their mass uncertainties into their model fitting procedures. They find a constant $M^*=10.78$ and mildly evolving values between 10.8 and 10.9. The work conducted in \citet{Leja2020} is an example of a study that included redshift evolution terms within a model fitting procedure that did not bin by redshift. Such a methodology can result in a smoothing of the evolution of the model parameters, which has its pros and cons depending on the timescales examined and assumptions made (e.g. fixing the slope of $\alpha_1$). For the propose of this study, we have elected to assume no particular evolutionary form for the GSMF.

These past studies, combined with our work trialling different functional forms to account for Eddington bias, suggest that up to 0.15~dex in variation of the characteristic mass $M^*$ can be attributed to implementing different methodologies. While significant enough to cause changes of a few $\sigma$, it is not enough to explain the $\sim0.5$~dex discrepancy among $M^*$ values found between these studies, indicating that there are further systematics at play between the data sets.

\section{Discussion}

\subsection{Time evolution of the high-mass component of the GSMF}

Including \textit{Spitzer}/IRAC photometry is found to be important in obtaining accurate galaxy classification and photometric redshifts, we therefore discuss the redshift evolution in the context of the GSMFs that make use of these data. We show the time evolution for the best-fit Schechter parameters ($\Phi_1, M^*, \alpha_1, \alpha_2$), corrected for Eddington bias via our non-parametric method, in Fig.~\ref{fig:Mstar}. Examining these best-fit Schechter parameters reveals evolution to be initially driven by the normalisation ($\Phi_1$) of the double Schechter components, which increases with time. This evolution is found to be much stronger at higher redshifts, with a change of over 0.5~dex between $z = 2$ through to $z = 1.25$. For $z<1.25$, the normalisation $\Phi_1$ stabilises and the evolution of the mass function becomes dominated by evolution in $M^*$ (there is also a possible flattening of $\alpha_1$, but this is degenerate with $\alpha_2$ which is poorly constrained at higher redshift).  

Under the assumption that $M^*$ is constant for $0<z<2$, the best fitting value of $\log_{10}(M^*/M_\odot) = 10.73\pm0.04$ is found for both SExtractor and ProFound derived photometry. However, we find that a constant $M^*$ model gives a poor quality of fit, with $\chi^2_{\rm red} = 3.4 \, (3.4)$ for SExtractor (ProFound). The observed rise in the value of $M^*$ at $z<1.25$ is present in all versions of the intrinsic GSMF that we produce with the differing Eddington corrections.
We therefore fit a simple linear model to the evolution of $M^*$ of the form $\log_{10}(M^*/M_\odot) = Az + B$, where $A$ and $B$ are free parameters and fit using minimisation of $\chi^2$. Introducing this simple time dependence into $M^*$ produces significantly better fits ($\chi^2_{\rm red} = 1.4 \, (2.4)$ with a Pearson correlation of 0.7 (0.75) for SExtractor (ProFound)) . The evolutionary fits can be described by,

\begin{equation}
    \frac{\Delta \log_{10}(M^*/M_\odot)}{\Delta z} = \left\{
    \begin{array}{l l}
      -0.164\pm0.047 & \text{SExtractor} \\
      -0.113\pm0.054 & \text{ProFound}\\
    \end{array} \right.
\end{equation}
a $2\text{--}3\sigma$ disagreement with the evolution described by a constant $M^*$.

However, the observed evolution in $M^*$ could also be attributed to a steeper evolution over a smaller redshift range, rather over the entire redshift range probed (see top-right panel of Fig.~\ref{fig:Mstar}). Such an evolution could be indicative of a break in the balance of growth channels that maintains a constant $M^*$ above $z\sim 1$ while $\Phi_1$ increases. Two such channels include star formation and the rate of major mergers. Observations and simulations show that the contribution towards new stellar mass from both mechanisms falls at $z<1.5$ in massive galaxies \citep{Rodriguez2015,Tomczak2016,Qu2017,Oleary2020}. The natural consequence of this would be to stabilise the high-mass end of the GSMF towards lower redshifts, as we find (e.g. the bottom right panel of Fig.~\ref{fig:MF} for $z<0.75$).

\subsection{The evolving quenched fraction of galaxies}

The effects of the evolving in situ star formation rate and merger rates of galaxies results in the changing shape of the GSMF. In order to probe this further, we examine the population of massive galaxies over the redshift range $0.5<z<1.5$ in more detail. We separate our sample of massive galaxies into two broad specific star formation rate (SSFR) bins, based on the same SED template used to derive the stellar mass. We term these bins `Star-forming' and `Quenched', which are defined as bins of $-8.5 > \log_{10}(SSFR / {\rm yr}^{-1}) > -10.0$ and $-10.0 > \log_{10}(SSFR/ {\rm yr}^{-1})$ respectively. The star forming bin is selected to cover the majority ($\sim95$ per cent) of the star forming main sequence at $z\sim1.5$ as found in both observational and simulation based studies \citep[e.g.][]{Sparre2015,Tomczak2016}. The quenched galaxies are thus defined as falling below this star forming main sequence. We do not enforce our definitions to be redshift dependent (i.e. a certain level above or below the evolving star-forming main sequence) to capture the global fall in star formation post-cosmic noon and produce results similar to those that use rest-frame colour selection \citep[e.g.][]{Davidzon17,McLeod2020}. We restrict the redshift range to $0.5<z<1.5$, as  SSFR estimates for galaxies at redshifts $z<0.5$ would become increasingly unreliable as the rest-frame UV exits the wavelength coverage of our observations.

\begin{figure} 
    \centering
    \includegraphics[width=\columnwidth]{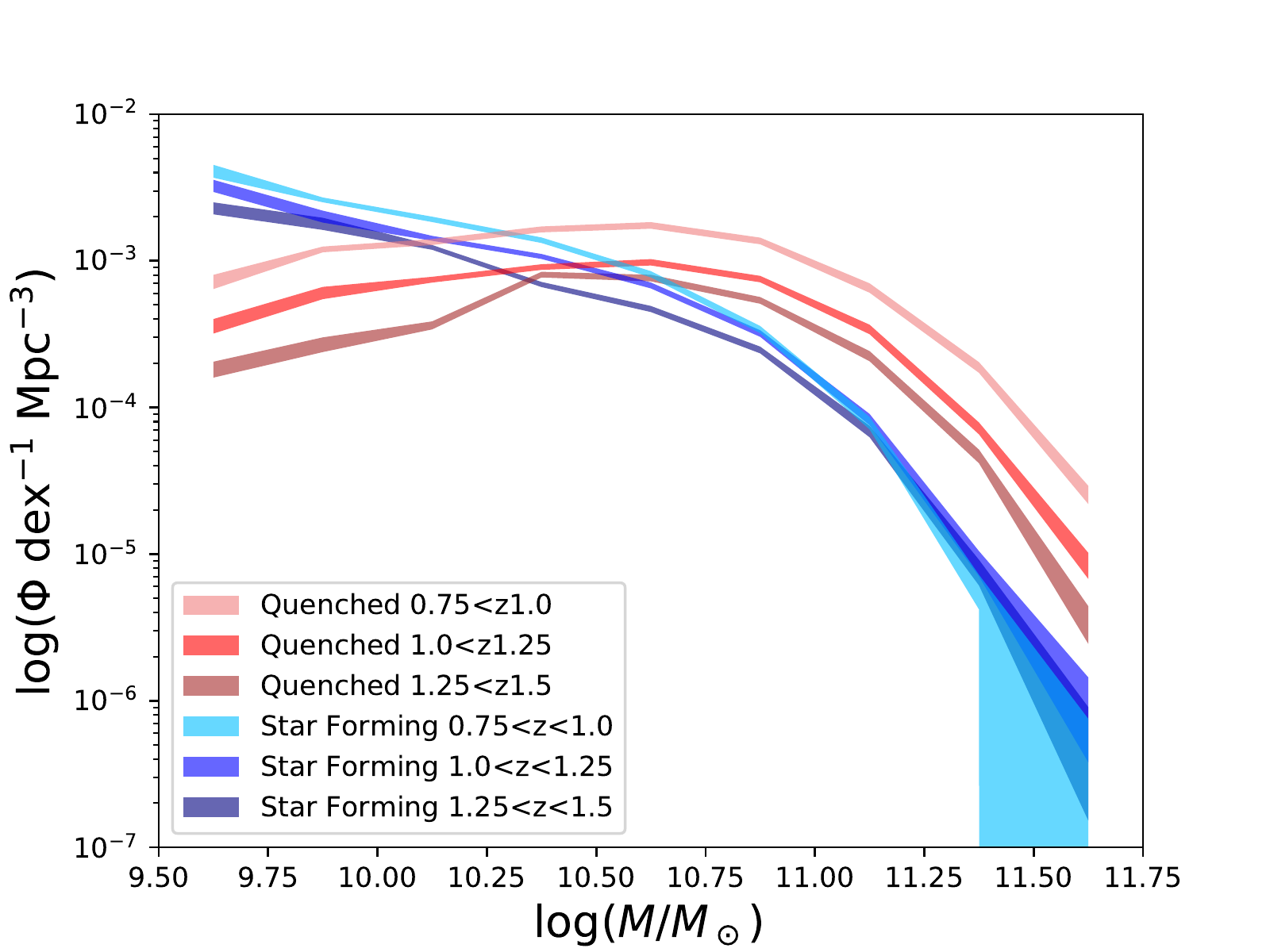}
    \caption{The GSMF when broken down into star forming and passive components by SSFR. The quenched population is shown with the red lines and the blue lines show the star-forming population. Darker shading indicates mass functions at higher redshifts. Shading widths indicates observational errors through both cosmic variance and $V_{max}$. The bin $0.5<z<0.75$ is not shown to reduce clutter and is found to be near identical to the bin $0.75<z<1.0$. The star-forming population is found to be near constant at higher masses while the passive population is found to evolve significantly between $1.5<z<0.75$.}
    \label{fig:PassiveMF}
\end{figure}

We reproduce the GSMF for these two populations in Fig.\ref{fig:PassiveMF}. Our definitions for star-forming and passive galaxies exclude a small number of galaxies undergoing starbursts $-8.5 < \log_{10}(SSFR/ {\rm yr}^{-1})$, and consequently the GSMFs in Fig.\ref{fig:PassiveMF} do not sum directly to the total GSMF presented in Fig.\ref{fig:MF}. We find the change in total number density of galaxies with $M>M^*$ to be driven by a strong evolution in the passive population. As a consequence of this trend, the most massive component of the GSMF becomes increasingly dominated by more passive galaxies while the number densities of star-forming galaxies are more constant, agreeing with \citet{Davidzon17} and \citet{McLeod2020}. Assuming a constant star formation rate, a borderline quenched galaxy with mass $M \geq M^*$, $\log_{10}(SSFR/ {\rm yr}^{-1}) =-10$ at $z=1.5$ would grow of order 0.15~dex in stellar mass during the $\sim4$~Gyr that passes between $0.5<z<1.5$. The majority of the objects in our high mass sample will produce fewer stars than predicted from this overly simple estimation, as most start with SSFR values which are lower and there is a global trend for the SSFR to decrease with time. Consequently, the growth of galaxies through solely main sequence evolution is insufficient to produce the observed increase in high-mass galaxy number densities. 

Since these quenched systems are not expected to form enough stars to move between bins of stellar mass ($\delta M\ll0.25$~dex), the two primary ways of increasing the quenched number densities at high masses are consequently merger events and the addition of newly quenched systems taken from the star-forming population. As our star-forming number densities at high mass remain near constant, this implies that these galaxies must be replenished by slightly lower mass star-forming galaxies at approximately the same rate at which they undergo quenching. Major mergers ($\sim1:4$ mass ratio) are capable of generating mass gains of order $\sim0.1$~dex for passive galaxies on top of internal star formation. Mergers themselves can trigger starbursts in the aftermath of the event. However, at later times ($z<1$), such mergers are increasingly `dryer' as massive galaxies tend to be gas poor \citep[e.g.][]{DeLucia2007,Lin2008,Davidzon2016,TOmczak2017}. Merger events come at the cost of reducing the total number of galaxies and its effects will be embedded in the evolving shape of the GSMF. In addition to providing pathways for stellar mass growth of passive galaxies, merger events can also be used to solve issues surrounding the radial size of passive galaxies. At higher redshifts ($z\sim1.5$), quenched galaxies are found to be compact relative to passive systems observed at $z=0$ \citep[e.g.][]{Williams2010,Wuyts2011,McLure2013}. The passive nature of these galaxies requires that external processes must be responsible for the observed shift in galaxy radius. The study by \citet{McLure2013} finds that combinations of major and minor mergers can simultaneously reproduce both the growth in stellar mass and radius observed in these systems between $z\sim1.5$ and today.

\begin{figure*}
    \centering
    \includegraphics[width=2\columnwidth]{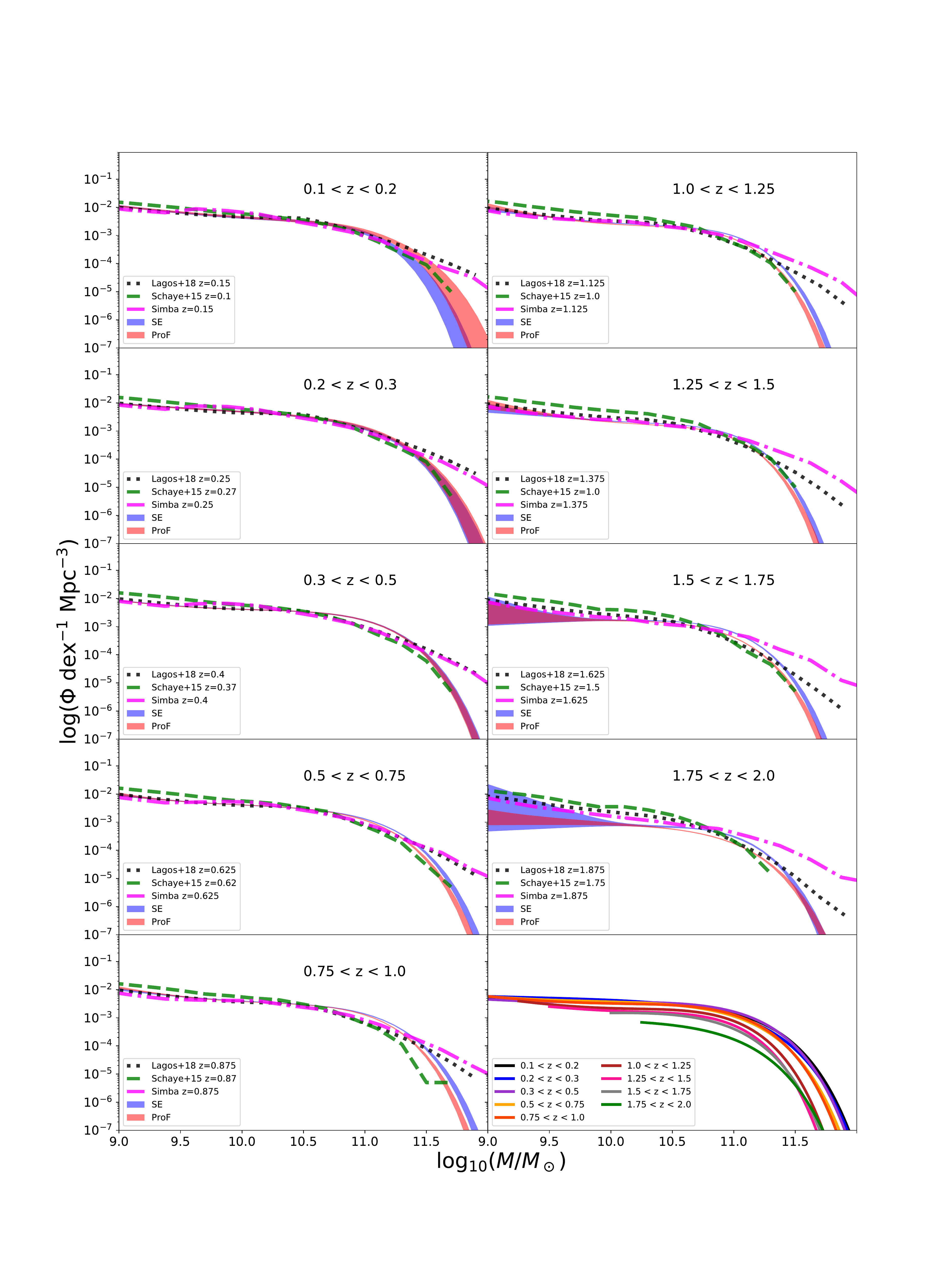}
    \vspace{-30pt}
    \caption{The intrinsic Galaxy Stellar Mass Function (GSMF) from Table~\ref{tab:Results} with Eddington bias removed. SExtractor derived results are in blue while ProFound is in red with the shading indicating the $1\sigma$ confidence interval. Presented alongside are examples of previously conducted simulations. For data from external sources, we present the closest redshift bin available for our comparisons. The studies used are the primary results from the \textit{SHARK} run conducted in \citet{Lagos18} as the black dotted lines, \textit{EAGLE} \citep{Schaye2015} as yellow dashed lines and \textit{Simba} as the dot/dashed purple lines \citep{Dave2019}.}
    \label{fig:Sims}
\end{figure*}

\subsection{Comparison to simulations}

In this section we compare our measured GSMF with the semi-analytic simulation (SAM) \textit{SHARK} \citep{Lagos18} and the hydrodynamical simulations \textit{Simba} \citep{Dave2019} and EAGLE \citep{Schaye2015}. In Fig.~\ref{fig:Sims} we show the GSMF results from these alongside our intrinsic (Eddington-bias corrected) double Schechter functions.
The \textit{SHARK}-SAM has the GSMF as one of physical measurements that it is tuned to reproduce at $z=0$, 1 and 2. As a result it is unsurprising to find excellent agreement with our results with the low mass slope and location of the primary `knee'. However, there is an excess in the number density of the most massive objects ($\log_{10}(M/M_\odot) > 11.5$) in the simulation. With the close matching to all other components of the GSMF, the likely source for this discrepancy could be in part the choice of GSMF which was used to  tune the simulation. The \citet{Lagos18} models elect to calibrate against studies such as \citet{Muzzin2013} and \citet{Wright18}, all studies with $M^*$ values on the higher end of the parameter space covered by observational studies ($10.8 < \log_{10}(M^*/M_\odot) > 11.0$). For our comparisons we utilise the GSMF derived from total masses provided from the primary SHARK run. The study by \citet{Lagos18} finds that implementing a 30kpc aperture to measure stellar mass leads to minimal impact for $\log_{10}(M/M_\odot) < 12.0$.

A similar situation is present when comparing against the results of the \textit{Simba} hydrodynamical simulation \citep{Dave2019}. \textit{Simba} implements a new, torque-limited accretion model \citep{Angles2017} for cold gas alongside more conventional Bondi accretion \citep{Bondi1952} for hot gas. The energy built up in accretion is used to fuel feedback and quench galaxies. While this model for black hole growth and AGN activity was considered to be more physically motivated over previous simulations, it is found to still over produce galaxies of very high stellar mass. Our observations reinforce the findings of \citet{Dave2019}, where the largest discrepancy in high mass galaxies is at $z\sim2$ (even when compared to the \textit{SHARK} results) and that there is a slight underproduction of galaxies around $M^*$ at low redshifts ($z<1$). This results in a much shallower `knee' than would be described with an exponential cut off. The authors explain possible causes of over-production of massive galaxies include over-merging of galaxies that blend, due to the use of a friend-of-friends (FoF) algorithm to count star particles, and the over-production of large halo masses.

The EAGLE simulation \citep{Schaye2015} implements AGN feedback through inputting a fraction of the accreted gas as thermal energy into the local surroundings. It is found to produce mass functions with slightly higher low-mass normalisation but much sharper cut offs at the high mass end. This results in an over production of galaxies on the low mass slope by a factor of around 2, but a closer match at $M \geq M^*$. The inconsistencies at the low-mass end are thought to originate from the lack of quenching of lower mass galaxies at higher redshifts, or from the need to implement more burst-like star formation histories \citep{Furlong2014}. These GSMF measurements implement three-dimensional apertures of radius 30kpc when calculating stellar masses, the size of which is found to reduce the mass assigned to the largest, most massive systems in the simulation and prevent some of the over-merging issues described in \citet{Dave2019}. This was found to reduce the number density of galaxies at $\log_{10}(M/M_\odot) \sim 11.5$ by around half a dex, imposing a steeper cut off in the number of high-mass systems. In the work conducted in \citet{Furlong2014}, the \textit{EAGLE} GSMF is fit with a double Schechter function and is found to exhibit a strong evolution in $M^*$ from $0.1<z<2.0$. The value of $\log_{10}(M^*/M_\odot) = 10.44 \pm 0.08$ at $z=2$ is notably small, even lower than the results from \citet{McLeod2020} which lie at the lower extrema of the range of observational constraints. The increase from $\log_{10}(M^*/M_\odot) = 10.74 \pm 0.05$ at $z=1$ to $\log_{10}(M^*/M_\odot) = 10.93 \pm 0.03$ at $z=0.1$ matches our ProFound-based GSMF to within $1\sigma$, while the value of $M^*$ is lower at $z=0.1$ from our SExtractor-derived GSMF. However, it is worth noting that \citet{Furlong2014} suggest that such an evolution in the characteristic mass could be the result of overly strong AGN feedback limiting the production of massive galaxies between $2<z<4$. It remains difficult to determine if faults lie within observations (mass errors, systematics etc.) or with the simulations (e.g. fine tuning AGN feedback) due to the sensitivity off all of these factors on the exponential decline in number density.

Within \textit{EAGLE}, the origin of this stellar mass growth in the most massive galaxies arises from multiple sources. \citet{Qu2017} show that in the redshift range of $z<1.5$, around 68 per cent of the massive galaxies ($\log_{10}(M/M_\odot)>11$) in \textit{EAGLE} undergo at least one major merger (at least 1:4 mass ratio, leading to $>0.1$~dex growth) and have the average fraction of stellar mass originating from outside the primary galaxy increase by a factor of two (from a median of 10 per cent to 20 per cent), with most of this external contribution fuelled by these major merger events. At $z=1$, over half of the simulated galaxies with $\log_{10}(M/M_\odot)>11$ are defined as quenched ($\log_{10}(SSFR/ {\rm yr}^{-1})<-10$), falling significantly below \textit{EAGLE} star-forming main sequence \citep{Furlong2014}. \ This leads to a similar fraction of quiescent galaxies to our sample. These consequently experience mass gains of less than 0.15~dex over the 4Gyr between $0.5<z<1.5$ under the simple assumption of a constant star formation rate. Creating this evolving value of $M^*$ while maintaining a near constant value of $\Phi_1$ thus likely requires a balance in merger events and internal star formation in order to sustain the growing number density of $M>M^*$ systems.

\section{Conclusions}

Utilising new photometric catalogues generated from optical and near-infrared data in the COSMOS and XMM-LSS fields, we have measured the GSMF over the redshift range of $0.1 < z < 2.0$, covering 60 per cent of the age of the Universe. The use of the two fields greatly reduces the impact of cosmic variance, due to both the wider area and the fact that they are widely separated on the sky, and allows for tight constraints on the GSMF at $M > M^*$. Simultaneously, the depth of the photometry available allows for constraints on on the GSMF of $\log_{10}(M/M_\odot) = 7.75$ for galaxies at $z \simeq 0.1$ and $\log_{10}(M/M_\odot) = 9.75$ for galaxies at $z\simeq 2.0$. We have investigated the use of ProFound and SExtractor source extraction software and also the impact of \textit{Spitzer}/IRAC [3.6] and [4.5] photometry on the GSMF. Our main conclusions are summarised as:

\begin{enumerate}
    \item The inclusion of \textit{Spitzer}/IRAC [3.6] and [4.5] photometry alleviates the degeneracies in select cases where the SEDs of red, massive galaxies in the redshift ranges $0.5<z<0.75$ and $1.0 < z < 1.25$ are confused with low redshift or stellar templates. Thus the inclusion of these data leads to fewer contaminants and a small decrease in the derived $M^*$ (of up to 0.1~dex) in these redshift ranges.
    \item Both SExtractor and Profound derived photometry produce consistent faint-end components of the GSMF. Differences between the mass functions at higher masses are greater when examining the extreme parts of our results (e.g. at the lowest and highest redshifts) but the resultant double Schechter fits are found to agree to within $\sim 1\sigma$.
    \item The measured GSMF is found to disagree with the assumption that the characteristic mass $M^*$ is constant with time between $0.1<z<2.0$ at the 3 (2) sigma level for the SExtractor (ProFound) derived results. Such an evolution in $M^*$ between $0.5<z<1.5$ can be seen in some (but not all) previous work and also in the predictions from the \textit{EAGLE} hydrodynamical simulation. However, significance is low and caveats in both the understanding of observational systematics and AGN feedback strength in simulations \citep{Furlong2014} mean a claim of an evolving $M^*$ is presently very mild.
    \item Eddington bias, and the methodology used to correct for it, is found to be highly influential on the shape of the high-mass end when attempting to retrieve the intrinsic GSMF from observations. There is presently still no consensus on the handling of observational and systematic errors that can impact stellar mass estimates \citep[see also discussions within][]{Grazain2015,Davidzon17}. For our data, we find the correction required to be asymmetric and poorly described by commonly used analytic forms. When comparing to the observed GSMF, applying a simple Gaussian treatment to the Eddington bias is found to reduce $M^*$ by around 0.05 dex. With the Lorentzian wings added into the description, the shift in $M^*$ doubles to around 0.1~dex. Utilising a non-parametric form, based on the measured error distribution, leads to $M^*$ shifts of 0.12~dex when compared to the observed GSMF.
    \item When splitting our galaxy sample by specific star formation rate (SSFR), our results confirm the findings of previous studies that show increasing number densities of quenched galaxies are responsible for the rise in the GSMF for $M>M^*$ \citep[e.g.][]{Davidzon17,McLeod2020}. Examining growth channels for stellar mass in the \textit{EAGLE} simulation show this to be the result of a combination of internal star formation and merger events. This is because internal star formation alone would amount to less than 0.15~dex in stellar mass growth for the majority of the population of massive galaxies ($M>M^*$). Between $0.5<z<1.5$, the constant number densities of star-forming galaxies at $M>M^*$ indicate that these galaxies quench at approximately the same rate that lower mass galaxies replace them through their own in situ star formation.
    \item Comparisons to simulations reveal that the semi-analytic model \textit{SHARK} and hydrodynamical simulation \textit{Simba} both over-produce massive galaxies at low and intermediate redshifts. Likewise the \textit{EAGLE} is found to over-predict the number of low-mass galaxies by a factor of around 2. This highlights that even in the era where such simulations can be fine tuned to better replicate observations, discrepancies still exist. It also further emphasises the conclusions drawn by these studies that additional work is required on all sides, from the observational data with which to calibrate/compare simulations against, to the development of the physical models used in simulations. However, we find the evolution of the high-mass component of the \textit{EAGLE} simulations replicates our observations of an evolving $M^*$.
\end{enumerate}

\section*{Acknowledgements}

The authors would like to pass on our thanks to J.Patterson and the University of Oxford's IT team at the Physics Department for their continued efforts. We give additional thanks to the HSC team for compiling spectroscopic catalogues from a vast range of surveys and granting easy public access to those catalogues. Our thanks are also offered to the referee of this publication, Dr Iary Davidzon, for the constructive feedback that was provided. This research  made  use  of Astropy,  a  community-developed core  Python  package  for  Astronomy  (Astropy  Collaboration,  2013).

NA acknowledges funding from the Science and Technology Facilities Council (STFC) Grant Code ST/R505006/1. This work was supported by the Glasstone Foundation, the Oxford Hintze Centre for Astrophysical Surveys which is funded through generous support from the Hintze Family Charitable Foundation and the award of the STFC consolidated grant (ST/N000919/1). 

This work is based on data products from observations made with ESO Telescopes at the La Silla Paranal Observatory under ESO programme ID 179.A-2005 and ID 179.A-2006 and on data products produced by CALET and the Cambridge Astronomy Survey Unit on behalf of the UltraVISTA and VIDEO consortia.

Based on observations obtained with MegaPrime/MegaCam, a joint project of CFHT and CEA/IRFU, at the Canada-France-Hawaii Telescope (CFHT) which is operated by the National Research Council (NRC) of Canada, the Institut National des Science de l'Univers of the Centre National de la Recherche Scientifique (CNRS) of France, and the University of Hawaii. This work is based in part on data products produced at Terapix available at the Canadian Astronomy Data Centre as part of the Canada-France-Hawaii Telescope Legacy Survey, a collaborative project of NRC and CNRS.

The Hyper Suprime-Cam (HSC) collaboration includes the astronomical communities of Japan and Taiwan, and Princeton University. The HSC instrumentation and software were developed by the National Astronomical Observatory of Japan (NAOJ), the Kavli Institute for the Physics and Mathematics of the Universe (Kavli IPMU), the University of Tokyo, the High Energy Accelerator Research Organization (KEK), the Academia Sinica Institute for Astronomy and Astrophysics in Taiwan (ASIAA), and Princeton University. Funding was contributed by the FIRST program from Japanese Cabinet Office, the Ministry of Education, Culture, Sports, Science and Technology (MEXT), the Japan Society for the Promotion of Science (JSPS), Japan Science and Technology Agency (JST), the Toray Science Foundation, NAOJ, Kavli IPMU, KEK, ASIAA, and Princeton University.

This paper makes use of software developed for the Large Synoptic Survey Telescope. We thank the LSST Project for making their code available as free software at  http://dm.lsst.org

This paper is based, in part, on data collected at the Subaru Telescope and retrieved from the HSC data archive system, which is operated by Subaru Telescope and Astronomy Data Center at National Astronomical Observatory of Japan. Data analysis was in part carried out with the cooperation of Center for Computational Astrophysics, National Astronomical Observatory of Japan.

\section*{Data Availability}
All imaging data was obtained from original sources in the public domain. Associated references to each survey utilised are provided within Section \ref{sec:data} and information on how to obtain the data are contained therein. A link to the online repository for the spectroscopic information used is also provided in Section \ref{sec:data}. The catalogues of photometry and photometric redshift estimations are expected to be released as part of a VIDEO data release later in 2021.

%%%%%%%%%%%%%%%%%%%%%%%%%%%%%%%%%%%%%%%%%%%%%%%%%%



%\bibliographystyle{mnras}

\bibliographystyle{mnras}
\bibliography{mnras_template} % if your bibtex file is called example.bib

%%%%%%%%%%%%%%%%%%%%%%%%%%%%%%%%%%%%%%%%%%%%%%%%%%

%%%%%%%%%%%%%%%%% APPENDICES %%%%%%%%%%%%%%%%%%%%%

\appendix
%\pagebreak
\section{Alternate double Schechter function fits}\label{alternatefits}

Displayed in Tables~\ref{tab:SE} and \ref{tab:PF} we show the results of conducting the MCMC fits to the raw data points as shown in \ref{fig:MF}. In Tables~\ref{tab:LogN} and \ref{tab:LxG} we show the results of the MCMC fits when utilising the first two Eddington bias corrections (single Gaussian, Gaussian multiplied by Lorentzian).

\begin{table*}
\caption{The median and standard deviations of the MCMC samples for the double Schechter function when using SExtractor based total photometry. Priors are set to ensure that $\alpha_1$ \& $\log_{10}(\Phi_1)$ refer to the high mass component and $\alpha_2$ \& $\log_{10}(\Phi_2)$ refer to the low mass component. Alongside are the number of galaxies from each of the two fields that contribute to the GSMF as shown in Fig.~\ref{fig:MF}. These mass functions are based on the raw observations and are not corrected for Eddington bias.}
\label{tab:SE}
\begin{tabular}{c|ccccc|cc}
\hline
Redshift & $\log_{10}(M^*/M_\odot)$ & $\alpha_1$ & $\log_{10}(\Phi_1)$ & $\alpha_2$ & $\log_{10}(\Phi_2)$ & COSMOS & XMM \\ 
SExtractor & & & $[\textrm{mag}^{-1} \textrm{Mpc}^{-3}]$ & & $[\textrm{mag}^{-1} \textrm{Mpc}^{-3}]$ & &  \\
\hline
0.1-0.2  &  $10.74^{+0.12}_{-0.12}$      & $-0.72^{+0.37}_{-0.31}$    & $-2.61^{+0.12}_{-0.15}$   & $-1.63^{+0.12}_{-0.18}$    &  $-3.48^{+0.31}_{-0.59}$ & 3099 & 4269  \\[4pt]
0.2-0.3  &  $10.95^{+0.06}_{-0.12}$      & $-1.24^{+0.46}_{-0.07}$    & $-2.89^{+0.11}_{-0.10}$   & $-1.61^{+0.29}_{-0.69}$    & $-5.31^{+2.02}_{-3.11}$ & 5423 & 7011  \\[4pt]
0.3-0.5  & $10.90^{+0.04}_{-0.04}$      & $-0.94^{+0.15}_{-0.10}$    & $-2.69^{+0.05}_{-0.05}$   & $-1.85^{+0.25}_{-0.34}$    & $-4.30^{+0.62}_{-0.86}$ & 15988 & 21292  \\[4pt]
0.5-0.75  &  $11.11^{+0.03}_{-0.03}$      & $-1.23^{+0.04}_{-0.03}$    & $-2.97^{+0.04}_{-0.05}$   & $-1.81^{+0.61}_{-0.77}$    & $-6.94^{+2.07}_{-2.04}$ & 26864 & 30713  \\[4pt]
0.75-1.0  &  $10.96^{+0.03}_{-0.03}$      & $-1.01^{+0.13}_{-0.07}$    & $-2.78^{+0.03}_{-0.04}$   & $-2.13^{+0.39}_{-0.49}$    & $-4.77^{+0.82}_{-1.01}$ & 33588 & 35190  \\[4pt]
1.0-1.25  &  $10.96^{+0.04}_{-0.04}$      & $-1.09^{+0.9}_{-0.09}$    & $-2.99^{+0.05}_{-0.06}$   & $-2.41^{+0.64}_{-0.41}$    & $-5.60^{+0.97}_{-0.83}$ & 25752 & 21395  \\[4pt]
1.25-1.5  &  $10.87^{+0.02}_{-0.02}$      & $-1.08^{+0.05}_{-0.04}$    & $-2.99^{+0.03}_{-0.04}$   & $-1.81^{+0.64}_{-0.74}$    & $-6.94^{+2.03}_{-2.08}$ & 18876 & 21575  \\[4pt]
1.5-1.75  &  $10.85^{+0.03}_{-0.03}$      & $-0.94^{+0.07}_{-0.06}$    & $-2.99^{+0.04}_{-0.05}$   & $-1.81^{+0.66}_{-0.74}$    & $-7.23^{+2.15}_{-1.89}$ & 13907 & 14130 \\[4pt]
1.75-2.0  &  $10.91^{+0.04}_{-0.04}$      & $-1.01^{+0.13}_{-0.10}$    & $-3.43^{+0.05}_{-0.07}$   & $-1.87^{+0.67}_{-0.70}$    & $-6.94^{+1.82}_{-2.08}$ & 9570 & 3778 \\ \hline
SExtractor+IRAC & & & & & \\
\hline
0.1-0.2  &  $10.83^{+0.10}_{-0.11}$      & $-0.72^{+0.38}_{-0.31}$    & $-2.73^{+0.11}_{-0.13}$   & $-1.60^{+0.06}_{-0.10}$    &  $-3.51^{+0.29}_{-0.55}$ & 3031 & 4264  \\[4pt]
0.2-0.3  &  $10.98^{+0.05}_{-0.08}$      & $-1.20^{+0.27}_{-0.06}$    & $-2.85^{+0.09}_{-0.09}$   & $-1.69^{+0.37}_{-0.70}$    & $-5.64^{+2.12}_{-2.77}$ & 5674 & 7239  \\[4pt]
0.3-0.5  &  $10.94^{+0.04}_{-0.04}$      & $-0.88^{+0.15}_{-0.11}$    & $-2.69^{+0.05}_{-0.05}$   & $-1.76^{+0.07}_{-0.09}$    & $-4.09^{+0.48}_{-0.63}$ & 15568 & 21246  \\[4pt]
0.5-0.75  &  $10.98^{+0.03}_{-0.03}$      & $-1.09^{+0.12}_{-0.07}$    & $-2.83^{+0.05}_{-0.05}$   & $-2.02^{+0.55}_{-0.59}$    & $-5.24^{+1.33}_{-1.61}$ & 25900 & 30209  \\[4pt]
0.75-1.0  &  $10.96^{+0.03}_{-0.03}$      & $-1.01^{+0.13}_{-0.07}$    & $-2.78^{+0.04}_{-0.04}$   & $-2.10^{+0.38}_{-0.48}$    & $-4.71^{+0.81}_{-1.00}$ & 31657 & 33791  \\[4pt]
1.0-1.25  &  $10.85^{+0.03}_{-0.03}$      & $-0.84^{+0.17}_{-0.09}$    & $-2.83^{+0.03}_{-0.04}$   & $-2.12^{+0.39}_{-0.48}$    & $-4.52^{+0.71}_{-0.76}$ & 25792 & 22288  \\[4pt]
1.25-1.5  &  $10.81^{+0.03}_{-0.03}$      & $-0.82^{+0.28}_{-0.13}$    & $-2.92^{+0.04}_{-0.08}$   & $-2.07^{+0.62}_{-0.58}$    & $-4.67^{+1.04}_{-0.90}$ & 18482 & 21133  \\[4pt]
1.5-1.75  &  $10.82^{+0.02}_{-0.02}$      & $-0.85^{+0.09}_{-0.06}$    & $-2.95^{+0.03}_{-0.04}$   & $-1.86^{+0.68}_{-0.71}$    & $-6.74^{+1.87}_{-2.22}$ & 13747 & 14020 \\[4pt]
1.75-2.0  &  $10.83^{+0.04}_{-0.04}$      & $-0.84^{+0.20}_{-0.12}$    & $-3.29^{+0.04}_{-0.06}$   & $-1.92^{+0.70}_{-0.68}$    & $-6.44^{+1.78}_{-2.42}$ & 9221 & 4476 \\[4pt] \hline
\end{tabular}
\end{table*}

\begin{table*}
\caption{The median and standard deviations of the MCMC samples for the double Schechter function when using ProFound based total photometry. Priors are set to ensure that $\alpha_1$ \& $\log_{10}(\Phi_1)$ refer to the high mass component and $\alpha_2$ \& $\log_{10}(\Phi_2)$ refer to the low mass component. Alongside are the number of galaxies from each of the two fields that contribute to the GSMF as shown in Fig.~\ref{fig:MF}. These mass functions are based on the raw observations and are not corrected for Eddington bias.}
\label{tab:PF}
\begin{tabular}{c|ccccc|cc}
\hline
Redshift & $\log_{10}(M^*/M_\odot)$& $\alpha_1$ & $\log_{10}(\Phi_1$) & $\alpha_2$ & $\log_{10}(\Phi_2)$ & COSMOS & XMM \\ 
ProFound & & & $[\textrm{mag}^{-1} \textrm{Mpc}^{-3}]$ & & $[\textrm{mag}^{-1} \textrm{Mpc}^{-3}]$ & &  \\
\hline
0.1-0.2  &  $10.88^{+0.23}_{-0.19}$      & $-1.06^{+0.42}_{-0.31}$    & $-2.81^{+0.21}_{-0.30}$   & $-1.76^{+0.20}_{-0.39}$    &  $-4.06^{+0.66}_{-2.29}$ & 3176 & 4062 \\[4pt]
0.2-0.3  &  $11.01^{+0.05}_{-0.07}$      & $-1.30^{+0.11}_{-0.05}$    & $-2.99^{+0.09}_{-0.09}$   & $-1.73^{+0.49}_{-0.72}$    &  $-6.47^{+2.52}_{-2.34}$ & 5384 & 6886 \\[4pt]
0.3-0.5  &  $10.83^{+0.04}_{-0.04}$      & $-0.90^{+0.16}_{-0.11}$    & $-2.66^{+0.05}_{-0.05}$   & $-1.88^{+0.22}_{-0.29}$    &  $-4.20^{+0.53}_{-0.68}$ & 16073 & 21130  \\[4pt]
0.5-0.75  &  $11.01^{+0.04}_{-0.04}$      & $-1.22^{+0.07}_{-0.04}$    & $-2.92^{+0.06}_{-0.06}$   & $-2.14^{+0.79}_{-0.62}$    &  $-6.23^{+1.50}_{-2.20}$ & 27059 & 38937  \\[4pt]
0.75-1.0  &  $10.86^{+0.03}_{-0.03}$      & $-0.94^{+0.13}_{-0.08}$    & $-2.72^{+0.03}_{-0.04}$   & $-2.11^{+0.26}_{-0.37}$    & $-4.45^{+0.57}_{-0.67}$ & 33820 & 33885 \\[4pt]
1.0-1.25  &  $10.85^{+0.03}_{-0.04}$      & $-0.92^{+0.15}_{-0.09}$    & $-2.89^{+0.04}_{-0.04}$   & $-2.35^{+0.37}_{-0.38}$    &  $-4.70^{+0.57}_{-0.65}$ & 25894 & 28041 \\[4pt]
1.25-1.5  &  $10.77^{+0.04}_{-0.04}$      & $-0.82^{+0.24}_{-0.12}$    & $-2.91^{+0.04}_{-0.05}$   & $-2.30^{+0.46}_{-0.47}$    &  $-4.54^{+0.71}_{-0.62}$ & 18959 & 14280 \\[4pt]
1.5-1.75  &  $10.78^{+0.03}_{-0.03}$      & $-0.84^{+0.11}_{-0.07}$    & $-2.89^{+0.03}_{-0.04}$   & $-1.90^{+0.70}_{-0.71}$    &  $-6.46^{+1.86}_{-2.41}$ & 13962 & 15248  \\[4pt]
1.75-2.0  &  $10.82^{+0.05}_{-0.06}$      & $-0.83^{+0.44}_{-0.16}$    & $-3.33^{+0.05}_{-0.08}$   & $-2.06^{+0.72}_{-0.60}$    &  $-5.49^{+1.33}_{-2.84}$ & 9595 & 4140 \\[4pt] \hline
ProFound+IRAC & & & & & & & \\
\hline
0.1-0.2  &  $10.99^{+0.14}_{-0.14}$      & $-0.96^{+0.42}_{-0.31}$    & $-2.83^{+0.16}_{-0.19}$   & $-1.66^{+0.16}_{-0.33}$    &  $-3.81^{+0.49}_{-1.52}$ & 2941 & 4120 \\[4pt]
0.2-0.3  &  $10.99^{+0.06}_{-0.08}$      & $-1.19^{+0.28}_{-0.07}$    & $-2.84^{+0.10}_{-0.09}$   & $-1.71^{+0.38}_{-0.70}$    &  $-5.45^{+1.98}_{-2.73}$ & 5600 & 7093 \\[4pt]
0.3-0.5  &  $10.93^{+0.04}_{-0.04}$      & $-0.89^{+0.15}_{-0.10}$    & $-2.68^{+0.05}_{-0.05}$   & $-1.89^{+0.15}_{-0.10}$    &  $-4.32^{+0.50}_{-0.70}$ & 15556 & 20820  \\[4pt]
0.5-0.75  &  $10.92^{+0.03}_{-0.03}$      & $-1.07^{+0.09}_{-0.07}$    & $-2.78^{+0.04}_{-0.05}$   & $-2.29^{+0.58}_{-0.48}$    &  $-5.38^{+1.08}_{-0.96}$ & 26014 & 38403 \\[4pt]
0.75-1.0  &  $10.89^{+0.03}_{-0.03}$      & $-0.94^{+0.12}_{-0.08}$    & $-2.74^{+0.03}_{-0.03}$   & $-2.17^{+0.30}_{-0.38}$    &  $-4.60^{+0.58}_{-0.75}$ & 31848 & 33188  \\[4pt]
1.0-1.25  &  $10.79^{+0.03}_{-0.03}$      & $-0.81^{+0.15}_{-0.10}$    & $-2.84^{+0.03}_{-0.04}$   & $-2.30^{+0.34}_{-0.40}$    &  $-4.62^{+0.59}_{-0.59}$ & 25860 & 27097  \\[4pt]
1.25-1.5  &  $10.77^{+0.04}_{-0.04}$      & $-0.80^{+0.26}_{-0.12}$    & $-2.96^{+0.04}_{-0.06}$   & $-2.21^{+0.44}_{-0.51}$    &  $-4.66^{+0.77}_{-0.59}$ & 18539 & 12523    \\[4pt]
1.5-1.75  &  $10.79^{+0.03}_{-0.03}$      & $-0.96^{+0.10}_{-0.07}$    & $-3.04^{+0.04}_{-0.05}$   & $-1.88^{+0.69}_{-0.68}$    &  $-6.77^{+1.77}_{-2.20}$ & 13785 & 11226  \\[4pt]
1.75-2.0  &  $10.88^{+0.04}_{-0.04}$      & $-1.17^{+0.14}_{-0.10}$    & $-3.52^{+0.06}_{-0.08}$   & $-1.86^{+0.65}_{-0.69}$    &  $-6.89^{+1.86}_{-2.11}$ & 9246 & 2898 \\[4pt] \hline
\end{tabular}
\end{table*}

\begin{table*}
\caption{The median and standard deviations of the MCMC samples for the double Schechter function when corrected for Eddington bias by convolving with a Gaussian distribution using a width of $\sigma_M=0.09$. Priors are set to ensure that $\alpha_1$ \& $\log_{10}(\Phi_1)$ refer to the high mass component and $\alpha_2$ \& $\log_{10}(\Phi_2)$ refer to the low mass component.}
\label{tab:LogN}
\begin{tabular}{c|ccccc}
\hline
Redshift & $\log_{10}(M^*/M_\odot)$& $\alpha_1$ & $\log_{10}(\Phi_1$) & $\alpha_2$ & $\log_{10}(\Phi_2)$  \\ 
SExtractor+IRAC & & & $[\textrm{mag}^{-1} \textrm{Mpc}^{-3}]$ & & $[\textrm{mag}^{-1} \textrm{Mpc}^{-3}]$   \\
\hline
0.1-0.2  &  $10.78^{+0.10}_{-0.11}$      & $-0.63^{+0.40}_{-0.33}$    & $-2.68^{+0.11}_{-0.13}$   & $-1.61^{+0.10}_{-0.14}$    &  $-3.44^{+0.26}_{-0.47}$  \\[4pt]
0.2-0.3  &  $10.95^{+0.06}_{-0.10}$      & $-1.17^{+0.39}_{-0.08}$    & $-2.82^{+0.10}_{-0.09}$   & $-1.65^{+0.32}_{-0.69}$    & $-5.07^{+1.19}_{-3.05}$   \\[4pt]
0.3-0.5  &  $10.90^{+0.04}_{-0.05}$      & $-0.79^{+0.18}_{-0.13}$    & $-2.66^{+0.05}_{-0.05}$   & $-1.69^{+0.17}_{-0.22}$    & $-3.86^{+0.39}_{-0.55}$   \\[4pt]
0.5-0.75  &  $10.93^{+0.04}_{-0.04}$      & $-1.01^{+0.18}_{-0.09}$    & $-2.78^{+0.05}_{-0.06}$   & $-1.94^{+0.44}_{-0.58}$    & $-4.60^{+1.03}_{-1.42}$   \\[4pt]
0.75-1.0  &  $10.91^{+0.03}_{-0.04}$      & $-0.92^{+0.18}_{-0.10}$    & $-2.74^{+0.04}_{-0.04}$   & $-1.93^{+0.31}_{-0.46}$    & $-4.28^{+0.67}_{-0.97}$   \\[4pt]
1.0-1.25  &  $10.79^{+0.03}_{-0.04}$      & $-0.68^{+0.24}_{-0.15}$    & $-2.79^{+0.03}_{-0.04}$   & $-1.90^{+0.31}_{-0.45}$    &  $-4.02^{+0.52}_{-0.76}$   \\[4pt]
1.25-1.5  &  $10.74^{+0.04}_{-0.05}$      & $-0.60^{+0.39}_{-0.20}$    & $-2.88^{+0.04}_{-0.11}$   & $-1.79^{+0.43}_{-0.67}$    &  $-3.94^{+0.63}_{-1.06}$   \\[4pt]
1.5-1.75  &  $10.77^{+0.02}_{-0.03}$      & $-0.78^{+0.11}_{-0.07}$    & $-2.91^{+0.03}_{-0.04}$   & $-1.85^{+0.66}_{-0.76}$    &  $-6.34^{+1.95}_{-2.48}$    \\[4pt]
1.75-2.0  &  $10.78^{+0.04}_{-0.05}$      & $-0.74^{+0.32}_{-0.14}$    & $-3.25^{+0.04}_{-0.06}$   & $-1.91^{+0.67}_{-0.71}$    &  $-5.91^{+1.67}_{-2.77}$   \\[4pt] \hline
ProFound+IRAC & & & & & \\
\hline
0.1-0.2  &  $10.94^{+0.15}_{-0.14}$      & $-0.88^{+0.45}_{-0.35}$    & $-2.79^{+0.15}_{-0.20}$   & $-1.64^{+0.14}_{-0.28}$    &  $-3.69^{+0.42}_{-1.27}$  \\[4pt]
0.2-0.3  &  $10.96^{+0.06}_{-0.10}$      & $-1.15^{+0.39}_{-0.09}$    & $-2.81^{+0.11}_{-0.10}$   & $-1.67^{+0.33}_{-0.69}$    &  $-4.94^{+1.66}_{-2.93}$  \\[4pt]
0.3-0.5  &  $10.89^{+0.04}_{-0.04}$      & $-0.81^{+0.16}_{-0.12}$    & $-2.64^{+0.04}_{-0.05}$   & $-1.80^{+0.20}_{-0.25}$    &  $-4.08^{+0.47}_{-0.62}$   \\[4pt]
0.5-0.75  &  $10.88^{+0.03}_{-0.03}$      & $-1.01^{+0.13}_{-0.07}$    & $-2.74^{+0.04}_{-0.05}$   & $-2.21^{+0.50}_{-0.50}$    &  $-4.97^{+1.02}_{-1.01}$  \\[4pt]
0.75-1.0  &  $10.84^{+0.03}_{-0.03}$      & $-0.83^{+0.15}_{-0.11}$    & $-2.70^{+0.03}_{-0.04}$   & $-1.99^{+0.24}_{-0.34}$    &  $-4.21^{+0.49}_{-0.66}$   \\[4pt]
1.0-1.25  &  $10.74^{+0.03}_{-0.04}$      & $-0.69^{+0.19}_{-0.13}$    & $-2.80^{+0.03}_{-0.03}$   & $-2.14^{+0.33}_{-0.41}$    &  $-4.28^{+0.51}_{-0.65}$   \\[4pt]
1.25-1.5  &  $10.71^{+0.03}_{-0.04}$      & $0.66^{+0.28}_{-0.15}$    & $-2.92^{+0.04}_{-0.05}$   & $-2.08^{+0.45}_{-0.54}$    &  $-4.19^{+0.61}_{-0.75}$     \\[4pt]
1.5-1.75  &  $10.74^{+0.03}_{-0.03}$      & $-0.90^{+0.10}_{-0.07}$    & $-2.99^{+0.04}_{-0.04}$   & $-1.84^{+0.65}_{-0.76}$    &  $-6.71^{+2.03}_{-2.23}$   \\[4pt]
1.75-2.0  &  $10.83^{+0.04}_{-0.04}$      & $-1.11^{+0.13}_{-0.11}$    & $-3.47^{+0.06}_{-0.07}$   & $-1.82^{+0.63}_{-0.76}$    &  $-6.96^{+2.05}_{-2.07}$  \\[4pt] \hline
\end{tabular}
\end{table*}

\begin{table*}
\caption{The median and standard deviations of the MCMC samples for the double Schechter function when corrected for Eddington bias through the scattering of photometry in the photo-z process, leading to a convolution with a Gaussian $\times$ Lorentzian functional form with a strength of $\sigma_M=0.50$. Priors are set to ensure that $\alpha_1$ \& $\log_{10}(\Phi_1)$ refer to the high mass component and $\alpha_2$ \& $\log_{10}(\Phi_2)$ refer to the low mass component.}
\label{tab:LxG}
\begin{tabular}{c|ccccc}
\hline
Redshift & $\log_{10}(M^*/M_\odot)$& $\alpha_1$ & $\log_{10}(\Phi_1$) & $\alpha_2$ & $\log_{10}(\Phi_2)$  \\ 
SExtractor+IRAC & & & $[\textrm{mag}^{-1} \textrm{Mpc}^{-3}]$ & & $[\textrm{mag}^{-1} \textrm{Mpc}^{-3}]$   \\
\hline
0.1-0.2  &  $10.73^{+0.11}_{-0.11}$      & $-0.53^{+0.42}_{-0.35}$    & $-2.64^{+0.10}_{-0.13}$   & $-1.60^{+0.10}_{-0.13}$    &  $-3.38^{+0.24}_{-0.40}$  \\[4pt]
0.2-0.3  &  $10.91^{+0.07}_{-0.12}$      & $-1.13^{+0.51}_{-0.11}$    & $-2.79^{+0.11}_{-0.10}$   & $-1.61^{+0.28}_{-0.66}$    & $-4.56^{+1.39}_{-3.21}$   \\[4pt]
0.3-0.5  &  $10.85^{+0.05}_{-0.05}$      & $-0.70^{+0.21}_{-0.16}$    & $-2.62^{+0.05}_{-0.05}$   & $-1.64^{+0.15}_{-0.20}$    & $-3.70^{+0.34}_{-0.49}$ \\[4pt]
0.5-0.75  &  $10.88^{+0.04}_{-0.05}$      & $-0.88^{+0.29}_{-0.15}$    & $-2.74^{+0.05}_{-0.06}$   & $-1.76^{+0.31}_{-0.51}$    & $3.98^{+0.68}_{-1.18}$   \\[4pt]
0.75-1.0  &  $10.85^{+0.04}_{-0.05}$      & $-0.75^{+0.25}_{-0.18}$    & $-2.70^{+0.04}_{-0.04}$   & $-1.72^{+0.22}_{-0.37}$    & $-3.76^{+0.44}_{-0.78}$   \\[4pt]
1.0-1.25  &  $10.70^{+0.05}_{-0.05}$      & $-0.33^{+0.29}_{-0.54}$    & $-2.77^{+0.04}_{-0.07}$   & $-1.62^{+0.20}_{-0.33}$    &  $-3.49^{+0.29}_{-0.54}$ \\[4pt]
1.25-1.5  &  $10.64^{+0.05}_{-0.05}$      & $-0.16^{+0.25}_{-0.77}$    & $-2.88^{+0.08}_{-0.11}$   & $-1.48^{+0.21}_{-0.56}$    &  $-3.38^{+0.25}_{-0.77}$  \\[4pt]
1.5-1.75  &  $10.69^{+0.04}_{-0.05}$      & $-0.58^{+0.36}_{-0.16}$    & $-2.86^{+0.04}_{-0.07}$   & $-1.78^{+0.59}_{-0.79}$    &  $-4.68^{+1.20}_{-2.97}$  \\[4pt]
1.75-2.0  &  $10.69^{+0.06}_{-0.08}$      & $-0.40^{+0.56}_{-0.33}$    & $-3.22^{+0.04}_{-0.09}$   & $-1.95^{+0.62}_{-0.68}$    &  $-4.57^{+0.82}_{-3.04}$  \\[4pt] \hline
ProFound+IRAC & & & & &  \\
\hline
0.1-0.2  &  $10.90^{+0.15}_{-0.15}$      & $-0.80^{+0.48}_{-0.38}$    & $-2.74^{+0.15}_{-0.20}$   & $-1.63^{+0.13}_{-0.25}$    &  $-3.60^{+0.37}_{-1.06}$  \\[4pt]
0.2-0.3  &  $10.91^{+0.08}_{-0.12}$      & $-1.08^{+0.52}_{-0.15}$    & $-2.78^{+0.12}_{-0.11}$   & $-1.62^{+0.28}_{-0.66}$    &  $-4.37^{1.21}_{-3.05}$  \\[4pt]
0.3-0.5  &  $10.85^{+0.05}_{-0.05}$      & $-0.73^{+0.20}_{-0.15}$    & $-2.60^{+0.05}_{-0.05}$   & $-1.73^{+0.18}_{-0.23}$    &  $-3.89^{+0.41}_{-0.56}$   \\[4pt]
0.5-0.75  &  $10.83^{+0.04}_{-0.05}$      & $-0.91^{+0.22}_{-0.10}$    & $-2.69^{+0.04}_{-0.05}$   & $-2.02^{+0.43}_{-0.54}$    &  $-4.39^{+0.87}_{-1.10}$  \\[4pt]
0.75-1.0  &  $10.77^{+0.04}_{-0.04}$      & $-0.63^{+0.21}_{-0.17}$    & $-2.65^{+0.03}_{-0.04}$   & $-1.80^{+0.19}_{-0.26}$    &  $-3.75^{+0.34}_{-0.52}$ \\[4pt]
1.0-1.25  &  $10.65^{+0.04}_{-0.05}$      & $-0.39^{+0.30}_{-0.56}$    & $-2.76^{+0.03}_{-0.05}$   & $-1.86^{+0.24}_{-0.33}$    &  $-3.74^{+0.35}_{-0.56}$  \\[4pt]
1.25-1.5  &  $10.60^{+0.05}_{-0.05}$      & $-0.23^{+0.31}_{-0.19}$    & $-2.90^{+0.05}_{-0.10}$   & $-1.70^{+0.27}_{-0.47}$    &  $-3.56^{+0.31}_{-0.63}$   \\[4pt]
1.5-1.75  &  $10.68^{+0.03}_{-0.05}$      & $-0.76^{+0.27}_{-0.11}$    & $-2.94^{+0.04}_{-0.05}$   & $-1.87^{+0.64}_{-0.73}$    &  $-5.44^{+1.58}_{-3.00}$   \\[4pt]
1.75-2.0  &  $10.78^{+0.05}_{-0.05}$      & $-1.02^{+0.18}_{-0.13}$    & $-3.41^{+0.06}_{-0.07}$   & $-1.85^{+0.64}_{-0.73}$    &  $-6.65^{+2.02}_{-2.26}$ \\[4pt] \hline
\end{tabular}
\end{table*}

%%%%%%%%%%%%%%%%%%%%%%%%%%%%%%%%%%%%%%%%%%%%%%%%%%

% Don't change these lines
\bsp	% typesetting comment
\label{lastpage}
\end{document}